\documentclass[conference]{IEEEtran}
\IEEEoverridecommandlockouts
\usepackage{cite}
\usepackage{amsmath,amssymb,amsfonts}
\usepackage{algorithmic}
\usepackage{graphicx}
\usepackage{textcomp}
\usepackage{xcolor}
\usepackage{caption}
\usepackage{subcaption}
\usepackage{listings}
\usepackage[
  bookmarks=true,
  bookmarksnumbered=true,
  draft=false,
  pagebackref=false,
  colorlinks=true,
  linkcolor=black,
  citecolor=black,
  filecolor=black,
  urlcolor=black,
  pdfpagelabels=false,
]{hyperref}
\usepackage{balance}

\usepackage{tikz}
\newcommand*\circled[1]{\tikz[baseline=(char.base)]{
            \node[shape=circle,draw,inner sep=0.5pt] (char) {#1};}}

\def\BibTeX{{\rm B\kern-.05em{\sc i\kern-.025em b}\kern-.08em
    T\kern-.1667em\lower.7ex\hbox{E}\kern-.125emX}}

\begin{document}
\pagestyle{plain}
\title{MicroGrad: A Centralized Framework for Workload Cloning and Stress Testing\\
\thanks{MicroGrad Tool: https://github.com/rgokulsm/MicroGrad}
}

\author{\IEEEauthorblockN{Gokul Subramanian Ravi}
\IEEEauthorblockA{\textit{ECE, UW-Madison} \\
rgokulsm@gmail.com}
\and
\IEEEauthorblockN{Ramon Bertran}
\IEEEauthorblockA{\textit{IBM Research} \\
rbertra@us.ibm.com}
\and
\IEEEauthorblockN{Pradip Bose}
\IEEEauthorblockA{\textit{IBM Research} \\
pbose@us.ibm.com}
\and
\IEEEauthorblockN{Mikko Lipasti}
\IEEEauthorblockA{\textit{ECE, UW-Madison} \\
mikko@engr.wisc.edu}
}

\maketitle

\begin{abstract}
We present \emph{MicroGrad}, a centralized automated framework that is able to efficiently analyze the  capabilities,  limits  and  sensitivities  of  complex  modern  processors  in  the  face  of  constantly  evolving application  domains.
\emph{MicroGrad} uses Microprobe, a flexible code generation framework as its back-end and a Gradient Descent based tuning mechanism to efficiently enable the evolution of the test cases to suit tasks such as Workload Cloning and Stress Testing.
\emph{MicroGrad} can interface with a variety of execution infrastructure such as performance/power simulators  as  well  as  native  hardware.
Further, the  modular ”abstract  workload  model”  approach to  building \emph{MicroGrad} allows  it  to  be  easily extended for further use.

In this paper, we evaluate \emph{MicroGrad} over different use cases and architectures and showcase that \emph{MicroGrad} can achieve greater than 99\% accuracy across different tasks within few tuning epochs and low resource requirements.
We also observe that \emph{MicroGrad's} accuracy is 25-30\% higher than competing techniques. 
At the same time, it is 1.5-2.5x faster or would consume 35-60\% less compute resources (depending on implementation) over alternate mechanisms.
Overall, \emph{MicroGrad}'s fast, resource efficient and accurate test case generation capability allow it to perform rapid evaluation of complex processors.
\end{abstract}

\section{Introduction}
\label{Introduction}

Analyzing the capabilities, limits and sensitivities of complex modern processors in the face of constantly evolving application domains is arduous and time consuming. 
Intelligently generating test cases which can efficiently perform the above analyses will enable quick turnaround times, thereby accelerating the final third of the Innovate-Build-Analyze cycle.

We are particularly interested in two challenging tasks under this umbrella of test case generation:
\begin{enumerate}
    \item \emph{Workload Cloning:} which extracts key execution characteristics of a real world application and models them into a synthetic workload.
    \item \emph{Stress Testing:}  that maximizes micro-architectural activity of a given processor, specifically to achieve worst-case estimates of execution metrics like performance and power.
\end{enumerate}

We present \emph{MicroGrad}, an open-source centralized automated framework that is able to efficiently analyze processors based on the scenarios described above.
\emph{MicroGrad} derives its name from \circled{1}\ \emph{Microprobe~\cite{microprobe}:} a flexible code generation framework which forms \emph{MicroGrad}'s back-end and \circled{2}\ \emph{Gradient Descent:} which is the tuning mechanism used by \emph{MicroGrad} to efficiently enable the evolution of the test cases to suit the tasks described earlier.

In the past, there has been considerable work in the domains of workload cloning~\cite{clone-1,clone-2,clone-3,clone-4,clone-5,clone-6} and stress testing~\cite{stress-1,stress-2,stress-3,microprobe,gest}. 
Despite this, open source frameworks for these goals have been scarce.
Meanwhile, the need for these tools is rapidly increasing with the momentum 
for open source hardware.
As the open source space grows, we will require systematic tools to characterize and stress-test the abundant varied designs and implementations.

\begin{figure}[thp]
\fbox{\includegraphics[width=\columnwidth,trim={0cm 0cm 0cm 0cm},clip]{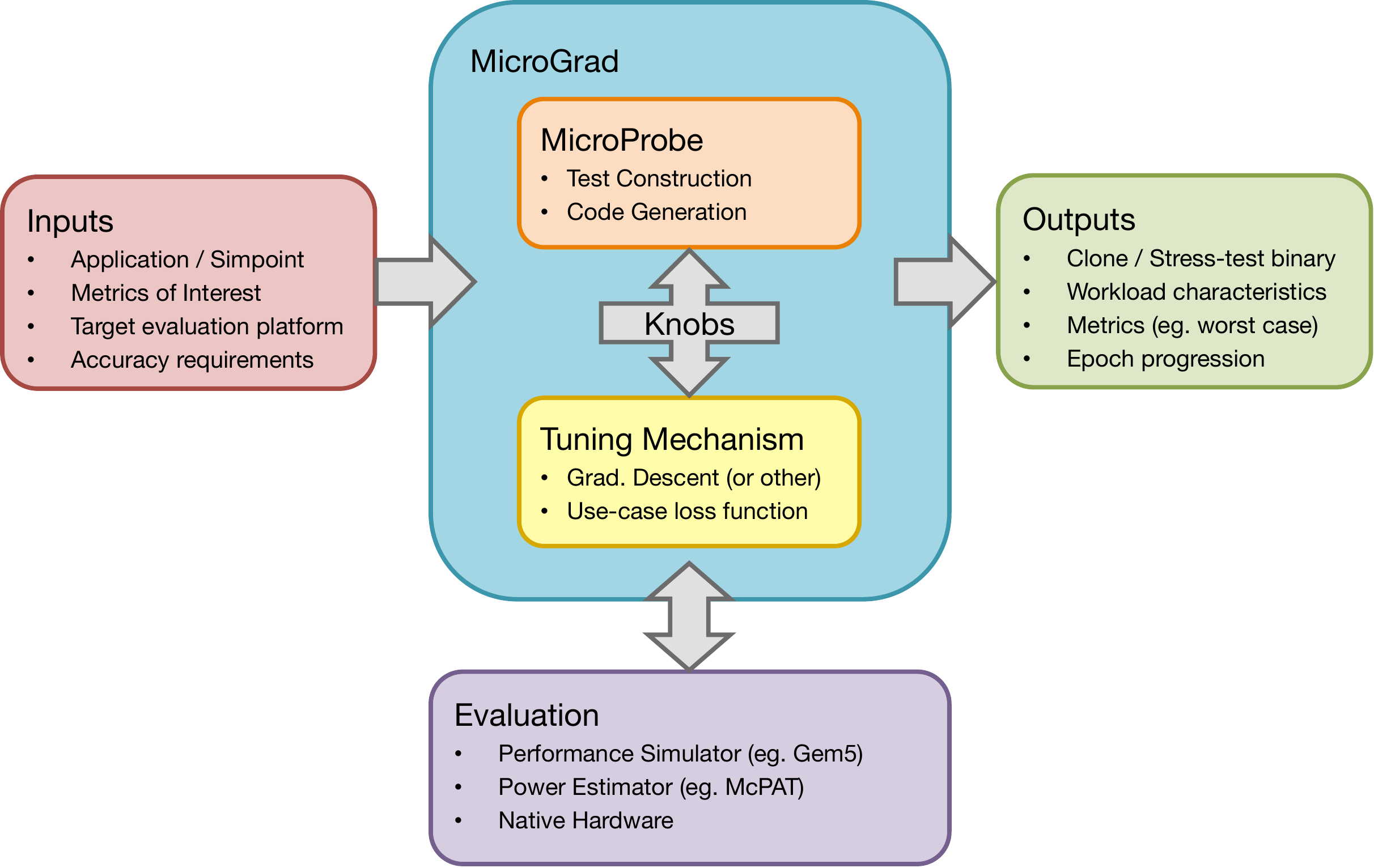}}
\centering
\caption{MicroGrad Overview}
\label{Overview_Fig}
\end{figure}

To our knowledge, two open-source frameworks available in this space are: Microprobe~\cite{microprobe} which can generate user-defined test cases, and GeST~\cite{gest} which uses Genetic Algorithm (GA) based evolution on an instruction-level model to generate stress tests.
\emph{MicroGrad} goes above and beyond the capabilities and use cases of the above, by providing a fast automated framework for a variety of purposes, all generated with a common centralized tuning mechanism and code generation back-end. 
Further, \emph{MicroGrad} can interface with a variety of execution platforms such as performance/power simulators as well as native hardware, in order to evaluate the processor architecture's execution efficiency.
Importantly, the modular "abstract workload model" approach to building \emph{MicroGrad} allows it to be easily developed upon - allowing for new use cases, improved tuning algorithms, as well as easy interfacing with new execution hardware and simulators.
An overview of \emph{MicroGrad} is shown in Fig.\ref{Overview_Fig}.

To the best of our knowledge, all prior approaches to cloning and stress testing have been either GA-based or expert driven.
Thus, the \emph{Gradient Descent} based tuning mechanism is a key novelty and highlight in the \emph{MicroGrad} framework.
The tuning mechanism is implemented over a gradient descent algorithm, which iterates through a sequence of "workload generation knobs" configurations (i.e. inputs to the \emph{Microprobe} framework) and evaluates a specified processor execution metric for those configurations. 
It gradually moves the code generation configuration in the direction of the steepest execution metric gradient, i.e. one which achieves the best metric improvement for every step change in the configuration, until the optimum configuration / convergence is reached.
Note that the execution metric is dependent on the use case - it could be a single high-level statistic like IPC or power consumption in the case of Stress Testing or a combination of both high-level and low-level statistics such as branch mispredictions, cache miss rates and IPC for Workload Cloning.
Overall, the tuning mechanism allows for fast and efficient convergence to the prescribed goal and is observed to considerably outperform competing tuning approaches.
Moreover, with its abstracted model, it is easier to deploy compared to expert-driven approaches.

\textbf{Summary of contributions:}
\begin{enumerate}
    \item We present \emph{MicroGrad}, an open-source automated framework for workload cloning and stress testing.
    To our knowledge, \emph{MicroGrad} is the first open tool for automated cloning and further, the only open tool for fast-exploratory stress testing with an abstract workload model.  
    \item \emph{MicroGrad} is the first proposal to perform intelligent test generation via a Gradient Descent based tuning mechanism, which is shown to outperform other tuning mechanisms and is easier to deploy than expert-driven approaches.
    \item \emph{MicroGrad} extends the potential for the Microprobe framework which has a wealth of features for code generation.
    \item The modular and abstracted approach to building \emph{MicroGrad} allows the seamless integration of new use cases, execution platforms and tuning algorithms.
\end{enumerate}
\section{Background}
\label{Background}

\subsection{Workload Cloning}
There are multiple challenges with using real-world applications for architecture benchmarking, such as intellectual property hurdles, effort involved in porting the application to suit the execution framework, as well as long run times.
While the advent of standardized benchmark suites have improved the testing ecosystem, there are still several time/resource challenges especially posed to architecture simulation in academic research as well as industry product development.
Simulation times are often intractable, even on today's most efficient simulators running on the fastest processing systems.

Workload Cloning is a general technique to mimic real-world applications or benchmarks via miniature synthetic workloads and has been pursued in multiple prior works~\cite{clone-1,clone-2,clone-3,clone-4,clone-5,clone-6}.
The technique distills key behavioral characteristics of the original application/benchmark and models them into a synthetic workload.
The resultant workload abstracts away any proprietary application characteristics, it is usually significantly shorter in execution time and it can be suitable compiled to make it amenable to both native hardware as well as simulation frameworks.
The integral components of the Cloning workflow are discussed below.

\subsubsection{Application Characteristics}
Specific characteristics of an application are captured and used to generate the synthetic workload.
These are characteristics which influence the instruction distribution, control flow as well as memory patterns of the application.
These characteristics can be divided into microarchitecture-independent and microarchitecture-dependent.
The former includes instruction distributions, register dependency distance etc. and memory footprints while the latter includes branch
misprediction and cache miss rates and others.
While some prior works~\cite{clone-1} have used both microarchitecture dependent and independent characteristics in conjunction, others have used a wider range of solely microarchitecture independent characteristics~\cite{clone-4}, but which are then significantly impacted by compiler optimizations.
In this work we use the former i.e. a combination of both microarchitecture dependent as well as independent characteristics which allow optimal capturing of both static and dynamic characteristics of an application on a specific processor architecture.

\subsubsection{Target Metrics}
The generated clones are expected to accurately meet specific target metrics.
A full system designer might require the clone to mimic the real application in terms of low-level target metrics such as L1/L2/TLB miss rates, branch misprediction rates, register usage, instruction distribution as well as high-level target metrics like IPC, power, energy or thermal characteristics. 
In this paper our tool evaluation focuses on cache miss rates, branch mispredictions, instruction distributions, IPC and Power.

\subsubsection{Generation Mechanism}
Prior clone generation mechanisms have comprised of a number of steps, each attempting to feed specific application/benchmark statistics into a model, so as to attempt to generate the required characteristics in the application. 
These steps include: generating the synthetic workload spine using instruction distribution, memory access pattern modeling, branch predictability modeling, register assignment, and finally code generation~\cite{clone-1}.
While these steps might individually achieve satisfactory accuracy for their low-level target metric (such as branch misprediction rate), performing them in a sequential manner (a sort of greedy approach) means that there is limited control over other high-level target metrics of interest like IPC.

In our work, we take a more synergic approach to clone generation. 
By estimating gradients and following the steepest curves, our tuning mechanism is able to inherently sacrifice the accuracy on some specific low-level target metric (for example, L2 cache miss rate) if required, if it aids in optimal achievement of other low-level and high-level target metrics, thereby creating a clone with higher fidelity.
Further, our approach allows a flexible generation time vs. cloning accuracy tradeoff. 
For instance, a 95\% accuracy 1-metric target would take considerably less generation time in comparison to a 99\% accuracy N-metric target.

\subsection{Stress testing}
Benchmark suites are usually built to represent the nominal behavior of real world applications and not to mimic worst-case scenarios.
However, worst-case scenarios in terms of microarchitectural activity, heat dissipation, power consumption and voltage noise~\cite{stress-1,stress-2,stress-3,stress-4,stress-5,stress-6} are critical to understand the limits and sensitivities of current generation processors, so that future systems can  migrate to the most promising regions of the microarchitectural design space.
These worst case scenarios are closely tied to the microarchitecture, and must be created in accordance.
Thus, stress tests are used to create these worst-case scenarios for a given target execution metric and a specific processor microarchitecture.
Considering the complexities and non-linear relationships within the modern processor, manually crafting such stress tests is usually time consuming and tedious. 
Consequently, automating their generation is of critical importance for a rapid design cycle.

\subsubsection{Generation Model}
As highlighted in prior work~\cite{gest}, there are two prominent design models for stress-test generation: a) based on
an abstract-workload model and b) based on instruction-level primitives.
In the abstract-model~\cite{stress-1,stress-2,stress-3} the stress test generation process involves tuning a vector of workload generation parameters/knobs such as instruction mix, register dependency distance, memory footprint / stride patterns and branch transition patterns. 
The vector is then used to generate the assembly (or high level language) code
On the other hand, for the instruction-level frameworks~\cite{gest,stress-4,stress-5}, the tuning is performed directly on the instruction assembly, with per-instruction control. 

The key advantage with the abstract workload model is that knobs are well defined, can be selected to be only a few in number, and can potentially have exclusive mapping to particular execution characteristics, significantly reducing the complexity of the tuning required to achieve the maximum stress. 
The advantage of the instruction-level model is that it provides deterministic and finer granularity of control i.e. on a per-instruction basis.
In this work, we adopt the abstract workload model, which provides suitable abstractions to allow for a more modular framework suited to multiple use cases, evaluation frameworks and tuning algorithms. 

\begin{table}[]
\centering
\resizebox{0.7\columnwidth}{!}{%
\begin{tabular}{|l|l|}
\hline
\textbf{Parameter}         & \textbf{Value}       \\ \hline
Population Size            & 50                   \\ \hline
Individual Size (\# knobs) & 25                   \\ \hline
Mutation Rate              & 3\%                  \\ \hline
Mutation position          & Random           \\ \hline
Mutation type              & Random  \\ \hline
Crossover Operator         & 1-point              \\ \hline
Crossover Rate             & 100\%                \\ \hline
Crossover Position         & Random           \\ \hline
Elitism                    & True                 \\ \hline
Tournament Size            & 5                    \\ \hline
\end{tabular}}
\caption{GA parameters}
\label{GA-param}
\end{table}

\subsubsection{Tuning Mechanism}
The role of the tuning mechanism is to nudge the generated test case towards maximum stress (as per the specified stress metric). 
Prior works built on both the generation models described above have predominantly utilized genetic algorithm (GA) based tuning.
GAs tune towards a target metric by applying operators inspired by natural evolution.
These operators include: selection of fittest individuals, crossover of features, mutation and guaranteed and elitism prioritization~\cite{stress-1,stress-2,stress-3, gest}.
The GA parameters used by prior work~\cite{gest} are shown in Table \ref{GA-param}.
To our knowledge, \emph{MicroGrad} is the only stress test generation scheme to stray away from GA based tuning.
We find that a gradient descent based tuning approach, with stochastic randomness to jump out of local minimas, as well as adaptive step sizes (larger to smaller over time), enable considerably faster (i.e. less number of tuning epochs) and more accurate convergence compared to the GA based approach.
For the abstract workload model specifically, our insight is that important GA operators like crossover are rather ineffective, while they are much more valuable in an instruction-level model. 
On the other hand, the gradient descent approach of following the steepest path to maximizing the metric of interest is very effective when local minimas can be avoided.

It is also interesting to note that the compute cost for a GD based tuning epoch is proportional to the number of knobs of interest, which could be low in the context of many use cases.
On the other hand, the compute cost in a GA epoch is proportional to the population size, which is often fixed throughout and therefore usually conservative.
Thus every GD epoch oj,is often faster and/or consumes less compute resources in comparison to the GA approach.
Our results demonstrate up to a 2.5x benefit for the GD approach.

\section{The MicroGrad framework}

An overview of the MicroGrad framework was shown in Fig.\ref{Overview_Fig}.
MicroGrad is built in a modular manner, allowing ease of use as well as flexibility for further development.
Additional use cases and metrics of interest, custom evaluation platforms, as well as improved tuning algorithms, can be developed and integrated conveniently into the framework.

\subsection{Framework Inputs}
The inputs to MicroGrad are provided in the form of a configuration file.
These inputs are use case dependent and those for our target use cases are described below.

\subsubsection{Workload Cloning}
\circled{1}\ The input specifies the target execution platform, the architecture configuration, the required cloning accuracy, as well as a maximum epoch limit for tuning.
If unspecified, defaults are used.
\circled{2}\ Further, characteristics of the application (which requires cloning) should be provided and there are multiple ways to do so:
\begin{itemize}
    \item The numerical values of the application's metrics of interest (which the clone is expected to match) can be directly provided as input. MicroGrad would then tune the clone to match these values.
    \item The application binary and its input data can be provided along with specification of the metrics of interest. By default, MicroGrad uses instruction distributions, cache miss rates, branch misprediction rates and IPC as the metrics of interest.
    \item Application Simpoints~\cite{simpoint} can be provided, so as to generate a clone for each simpoint individually. The combination of simpoints and clones can expand the evaluation space of the original application, with potentially one clone for each interesting phase of the application.
\end{itemize}

\subsubsection{Stress Testing}
\circled{1}\ The input specifies the target execution platform, the architecture configuration and a maximum epoch limit for tuning.
\circled{2}\ Metrics of stress are provided as inputs - this can either be a single high-level metric such as IPC or a single low-level metric like branch misprediction rate or a combination of multiple metrics.
By default, IPC is used as the stress metric.

\begin{lstlisting}[caption={Example Knobs and range of values},captionpos=b]
#Instruction fractions
ADD = [1, 2, 3, 4, 5, 6, 7, 8, 9, 10]
MUL = [1, 2, 3, 4, 5, 6, 7, 8, 9, 10]
FADDD = [1, 2, 3, 4, 5, 6, 7, 8, 9, 10]
FMULD = [1, 2, 3, 4, 5, 6, 7, 8, 9, 10]
BEQ = [1, 2, 3, 4, 5, 6, 7, 8, 9, 10]
BNE = [1, 2, 3, 4, 5, 6, 7, 8, 9, 10]
LD = [1, 2, 3, 4, 5, 6, 7, 8, 9, 10]
LW = [1, 2, 3, 4, 5, 6, 7, 8, 9, 10]
SD = [1, 2, 3, 4, 5, 6, 7, 8, 9, 10]
SW = [1, 2, 3, 4, 5, 6, 7, 8, 9, 10]

#Dependency distance
REG_DIST = [1, 2, 3, 4, 5, 6, 7, 8, 9, 10] 

#Memory Footprint
MEM_SIZE = [2, 4, 8, 16, 32, 64, 128, 256, 512, 1024, 2048]

#Memory acceess strides
MEM_STRIDE = [8, 12, 16, 20, 24, 32, 40, 48, 56, 64]

#Memory temporal locality - how many to repeat
MEM_TEMP1 = [1, 2, 4, 8, 16, 32, 64, 128, 256, 512]

#Memory temporal locality - how often to repeat
MEM_TEMP2 = [1, 2, 3, 4, 5, 6, 7, 8, 9, 10]

#Branch pattern randomization ratio
B_PATTERN = [0.1, 0.2, 0.3, 0.4, 0.5, 0.6, 0.7, 0.8, 0.9, 1]
\end{lstlisting}

\subsection{Knob Interface}
MicroGrad uses a set of knobs to interface between the Tuning mechanism and the Microprobe framework.
The Tuning mechanism nudges the knobs in the directions appropriate for the use case, and these knobs are conveyed to Microprobe which generates the test-case based on these knob values.
Further, the generated test case is executed on the evaluation framework, whose outputs metrics are fed back to the tuning mechanism to re-tune the knob values. 
An example subset of the knobs used by MicroGrad and their range of values are shown in Listing 1.
In this example subset, the instruction knobs act as fractions of the overall distribution, another knobs allows control of the register dependency distance, the memory knobs specify footprint, stride and temporal locality, and the branch pattern knob specifies the fraction of randomness in the branch pattern.
Other tuning knobs are not shown in the interest of space.


\begin{lstlisting}[caption={Microprobe passes},captionpos=b,escapeinside={(*}{*)}]
passes = [
    # Create a container with required size
    SimpleBuildingBlockPass(loop_size),

    # Reserve special registers
    ReserveRegistersPass(reserved_registers),

    # Set instruction profile
    SetInstructionTypeByProfilePass(PROFILE),

    # Initialize registers
    InitializeRegistersPass(value=RNDINT),

    # Randomize some branch directions
    RandomizeByTypePass(
        branch_instrs,  # to replace
        isa.instructions['BGE_V0'],
        BRANCH_RAND,  # randomize probability
        ),

    # Memory streams with footprint, stride pattern and ratio of accesses
    GenericMemoryStreamsPass(
        [[1, SIZE1, RATIO1, STRIDE1, 1, 0],
         [2, SIZE2, RATIO2, STRIDE2, 1, 0]]
        ),

    # Assign operands as per required dependency distance
    DefaultRegisterAllocationPass(dd=REG_DIST),

    # Check and update addresses
    UpdateInstructionAddressesPass()
]
\end{lstlisting}

\subsection{Code Generation}
The tuning mechanism presents knob values to Microprobe~\cite{microprobe,microprobeweb} to generate the corresponding test case.
Microprobe is a flexible code generation framework that provides a high level Python scripting interface to access to a rich set of mechanisms and features to control the code generation process. This enables the users to adapt the code generation process to different use cases without having to deal with all low level details. 
For instance, it has been used in the past for power model generation~\cite{microprobe}, maximum power and dI/dt stressmark generation~\cite{stress-6}, complete architecture characterizations~\cite{z13,power9} and also reliability analysis~\cite{ser}.    

MicroGrad directly uses Microprobe scripting interface to define the code generation process according the knobs specified.  
The test case is then generated by a sequence of code synthesis passes which are applied in accordance with the MicroGrad defined ordering rules. 
 A code snippet highlighting some of the standard Microprobe passes used by MicroGrad are shown in Listing 2. 
More details on these passes and others can be found on the open source Microprobe tool website~\cite{microprobeweb}.

\begin{lstlisting}[caption={Gradient descent  tuning},captionpos=b,escapeinside={(*}{*)}]
def eval(Target):
    """
    Tuning to reach Target
    """
    
    # Run to convergence / target
    # KC is knob configuration
    while True:
        itn++ #epoch iteration
            
        #initial or continuation
        if !KC: KC_base = random()
        else: KC_base = KC
        
        #KC via Microprobe + HW gives Metric
        Met_base = HW((*$\mu$*)(KC_base)) 
        
        #step size varies over epochs
        step_size = step_array(itn)
           
        #Perform epoch
        KC = epoch(KC_base, Met_base, Target, step_size)
        
        #Check for convergence or target
        if (KC - KC_base) < (*$\epsilon$*) 
           || (KC - Target) <  (*$\epsilon$*) :
                break
return KC

def epoch(kc, Met_base, Target, step_size):
    """
    GD to create new knob configuration
    """
    
    # Iterate over all knobs
    while not kc.finished:
        if not_skip: #not skipping this knob
                
            # Perturb ith knob
            kc_i = modify(kc, i, (*$\delta$*))
        
            # Calculate h/w metric at kc_i
            Met = HW((*$\mu$*)(kc_i))
            
            # Loss at kc_i
            Loss = L(Met, Met_base, Target)
            
            # Compute the partial derivative
            grad[i] = (Loss / step_size)
            
        # Step to next knob and reset current
        kc.iternext()

    # Calculate new configuration
    kc = kc - step_size*grad
            
return kc
\end{lstlisting}

\subsection{Gradient-based Tuning}

Each tuning epoch involves tuning the knob configuration by evaluating the execution metrics in the vicinity of the current configuration and making changes to the knobs accordingly.
Pseudo-code for the tuning mechanism is shown in Listing 3 and features of the mechanism are discussed below.

\circled{1}\ A new tuning epoch starts with capturing the execution metrics (eg. IPC, energy, cache miss rates) at the previous epoch's output knob configuration (random configuration, if first epoch).  
This is the 'base' configuration for this epoch.
This involves generating the test case with Microprobe at the base configuration, running the test case on the evaluation platform and measuring the base metrics.  

\circled{2}\ The goal at the end of the epoch is to find the new knob configuration which is the steepest move (in terms of the matching the use case requirements) from the base configuration.

\circled{3}\ In order to achieve this, the base knob configuration is independently perturbed by +/- $\delta$ in each dimension (i.e. each knob). Each resulting configuration is a 'gradient-check' configuration.
This results in 2*knobs number of 'gradient-checks' per epoch.

\circled{4}\ The execution metrics are then captured at each of these 'gradient-check' configurations, again by generating test cases with Microprobe and running the test cases on the evaluation platform.

\circled{5}\ For each case, the 'gradient-check' execution metrics are compared to the base and the target metrics to obtain a Loss, which is tied to the use case goal.

\circled{6}\ The gradient of the Loss along each knob dimension is calculated by evaluating how much the loss function changed along each dimension's $\delta$ perturbation. 

\circled{7}\ This information is used to obtain the new knob configuration - the knobs with the steepest gradients move by 'one' step-size, while the other knobs proportionally move by a fractional of the step size. 
This becomes the starting configuration for the next epoch.

\circled{8}\ Inspired by adaptive learning rate based gradient methods~\cite{adam}, the tuning mechanism's step-sizes are larger on earlier epochs and gradually become smaller, allowing for rapid convergence earlier but slower but surer convergence later on.

\circled{9}\ To add robustness to the convergence to help avoid local minima, a random set of knobs are skipped in tuning each iteration, with decreasing skipping probability over epochs. 

\circled{10}\ Tuning continues until either convergence, the target accuracy or the maximum number of epochs is reached.

\subsection{Metric Evaluation}
Once the test case is generated and compiled to meet the requirements of the evaluation architecture, the test case is executed on the platform.
MicroGrad is able to interface with a number of platforms such as native hardware, performance simulators (e.g. Gem5~\cite{gem5}) and power estimation frameworks (eg. McPAT~\cite{mcpat}).
In the case of simulators, the architecture configuration can be passed as input to MicroGrad and used in the simulator to express the desired architecture.

In terms of capturing metrics, the requisite metrics are dependent on the use case. 
A stress test use case might require only IPC / Power, whereas a cloning use case might require low-level metrics like mispredictions and miss rates.
When using simulators, the MicroGrad interface enables the required metrics to be read from the output dumps of the simulators. In the case of native hardware evaluation, appropriate hardware counters and their required interfacing can be used in similar fashion.

\subsection{Framework Outputs}
MicroGrad completes execution when either the target is met or some execution time/resource constraint is reached.
The output at the end of execution is dependent on the use case.
In the case of Workload Cloning, MicroGrad outputs the clone binary, details of the corresponding knobs and the metrics based on the evaluation of the clone.
With stress testing, MicroGrad outputs the stress test binary, the knobs and the stress metrics.
In both scenarios, intermediate data can be stored, so as to understand the tuning/execution progress over the epochs (for example, to improve the tuning algorithm).
\section{Evaluation}

\begin{table}[]
\centering
\resizebox{0.9\columnwidth}{!}{%
\begin{tabular}{|l|l|l|}
\hline
\textbf{Parameter}       & \textbf{Small} & \textbf{Large}       \\ \hline
\textbf{Frequency}       & \multicolumn{2}{l|}{2 GHz}            \\ \hline
\textbf{Front-End Width} & 3              & 8                    \\ \hline
\textbf{ROB/LSQ/RSE}     & 40/16/32       & 160/64/128           \\ \hline
\textbf{ALU/SIMD/FP}     & 3/2/2          & 6/4/4                \\ \hline
\textbf{L1/L2 Cache}     & 16k/256k     & 32k/1M + prefetch \\ \hline
\textbf{Memory}          & \multicolumn{2}{l|}{1GB}              \\ \hline
\end{tabular}%
}
\caption{Core Configuration}
\label{Core}
\end{table}

\begin{figure*}[t]
  \centering
    \begin{subfigure}{0.24\textwidth}
    \fbox{\includegraphics[width=\textwidth,trim={1.25cm 2cm 2.5cm 1.25cm},clip]{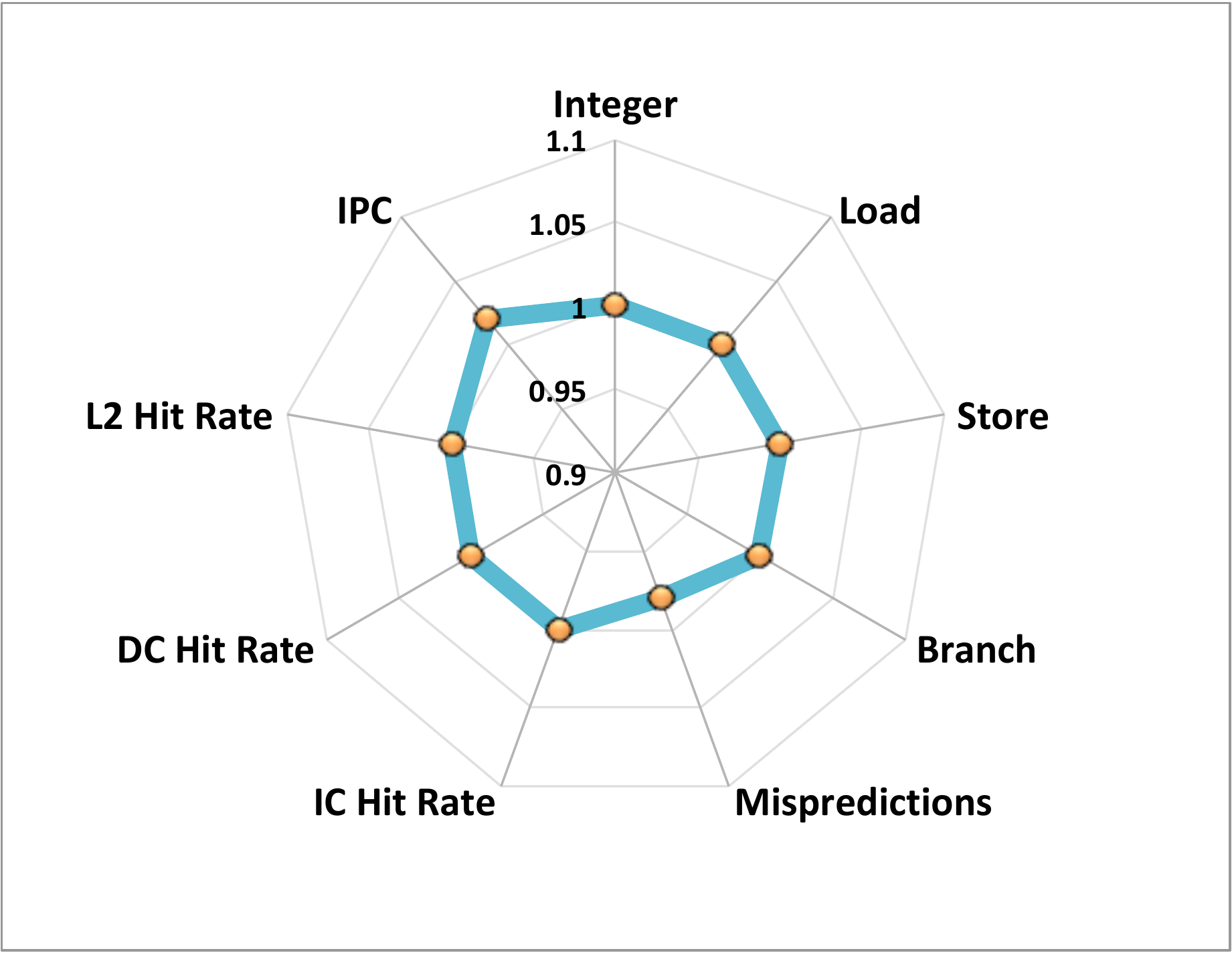}}
    \end{subfigure}
    \begin{subfigure}{0.24\textwidth}
    \fbox{\includegraphics[width=\textwidth,trim={1.25cm 2cm 2.5cm 1.25cm},clip]{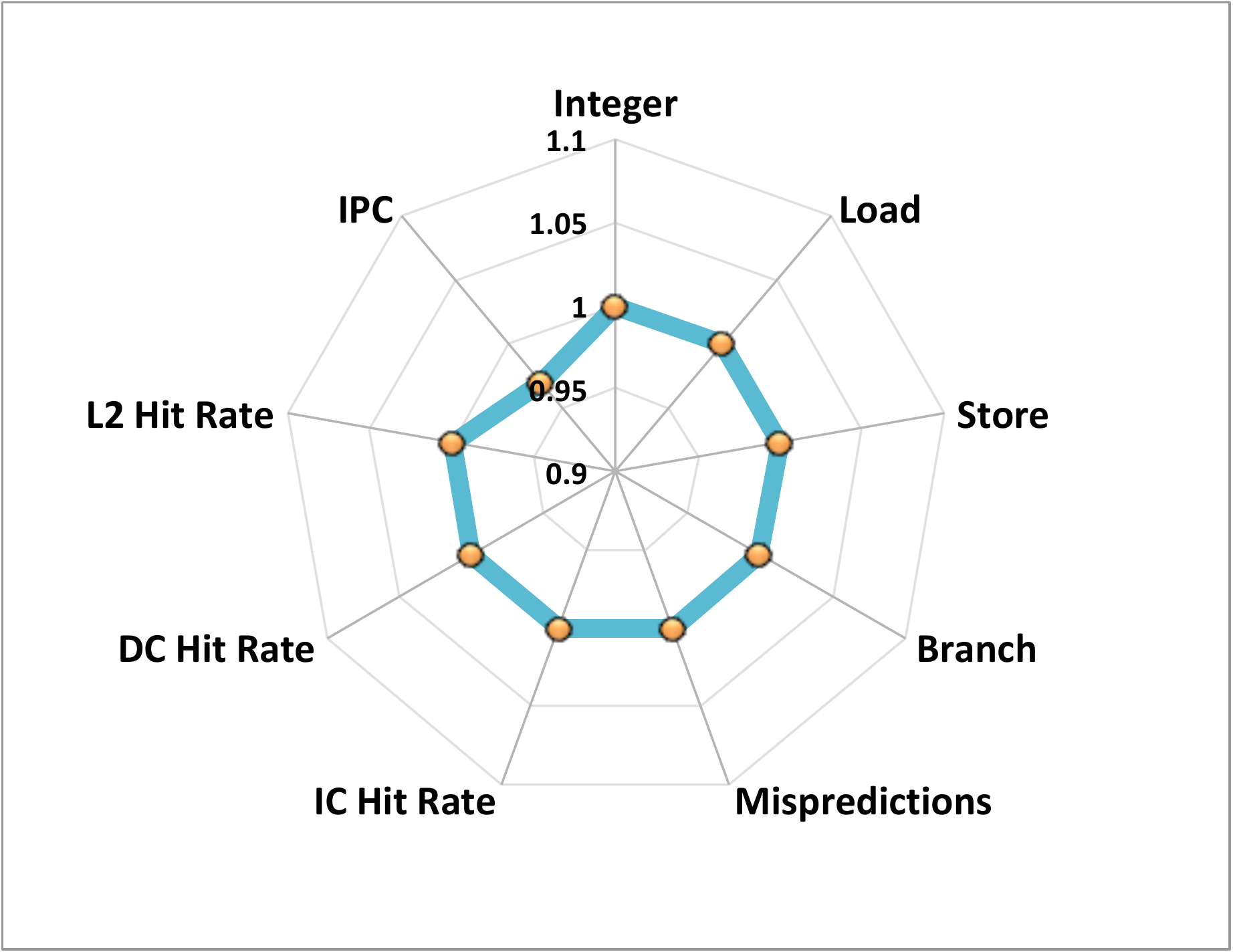}}
    \end{subfigure}
    \begin{subfigure}{0.24\textwidth}
    \fbox{\includegraphics[width=\textwidth,trim={1.25cm 2cm 2.5cm 1.25cm},clip]{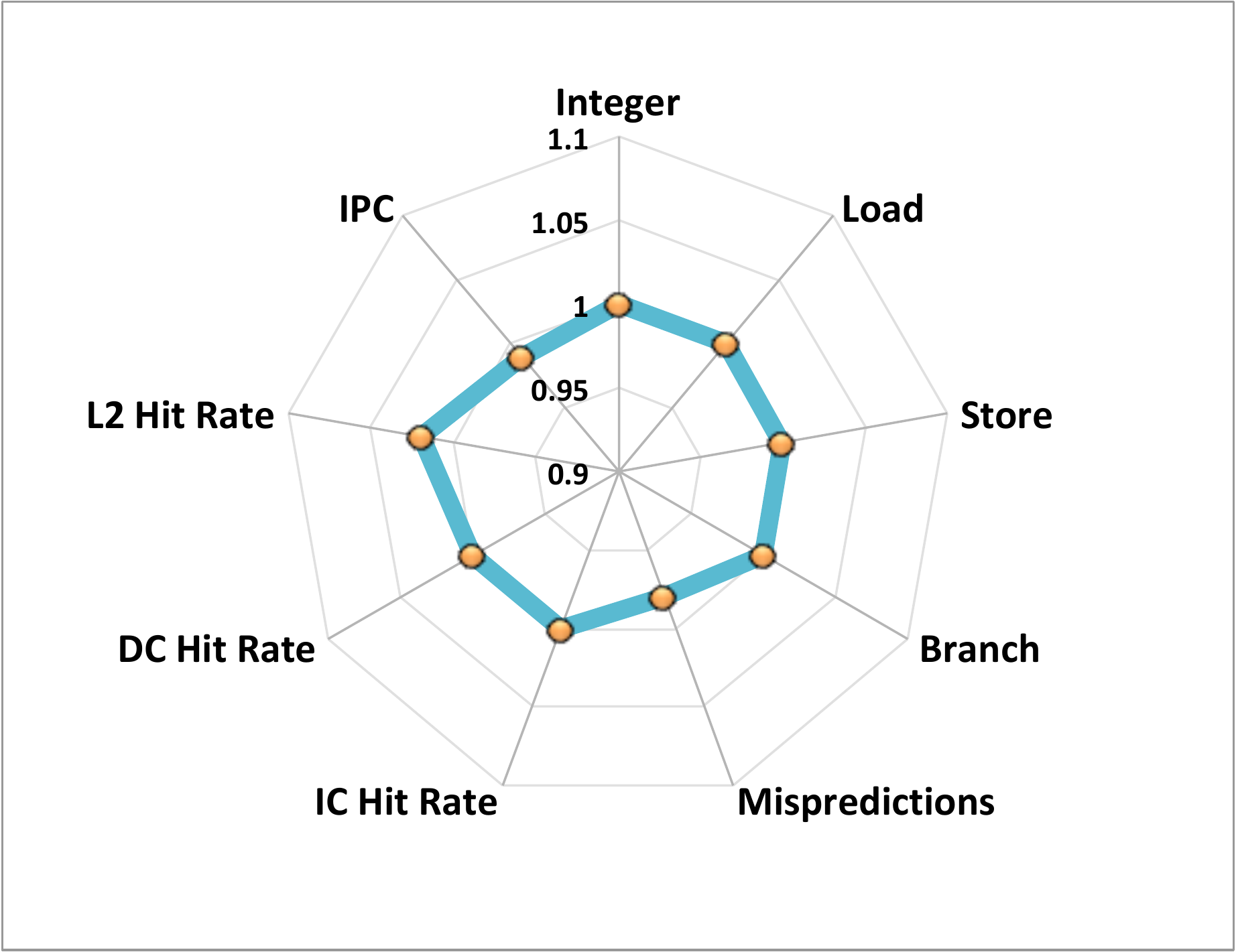}}
    \end{subfigure}
    \begin{subfigure}{0.24\textwidth}
    \fbox{\includegraphics[width=\textwidth,trim={1.25cm 2cm 2.5cm 1.25cm},clip]{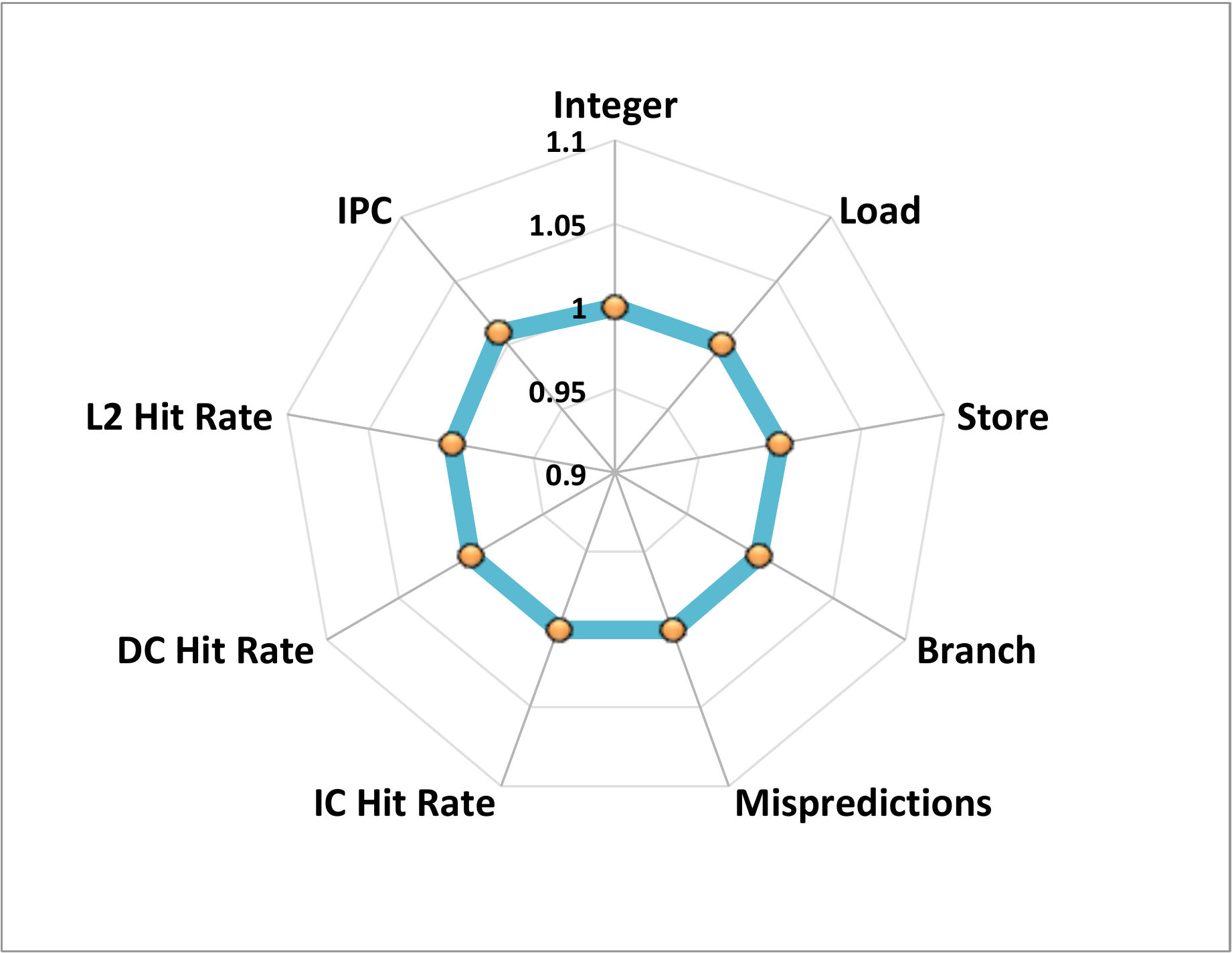}}
    \end{subfigure}
    \begin{subfigure}{0.24\textwidth}
    \fbox{\includegraphics[width=\textwidth,trim={1.25cm 2cm 2.5cm 1.25cm},clip]{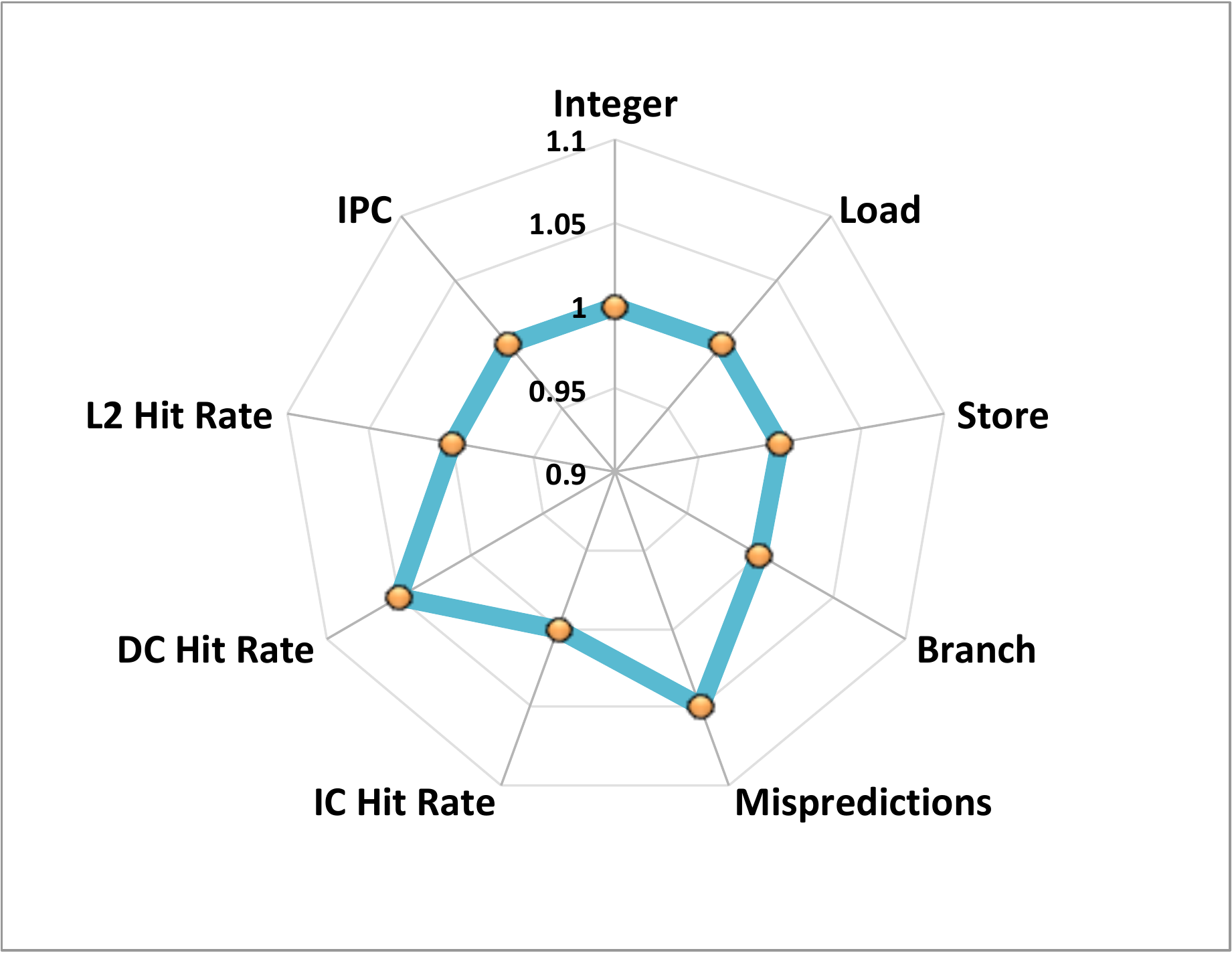}}
    \end{subfigure}
    \begin{subfigure}{0.24\textwidth}
    \fbox{\includegraphics[width=\textwidth,trim={1.25cm 2cm 2.5cm 1.25cm},clip]{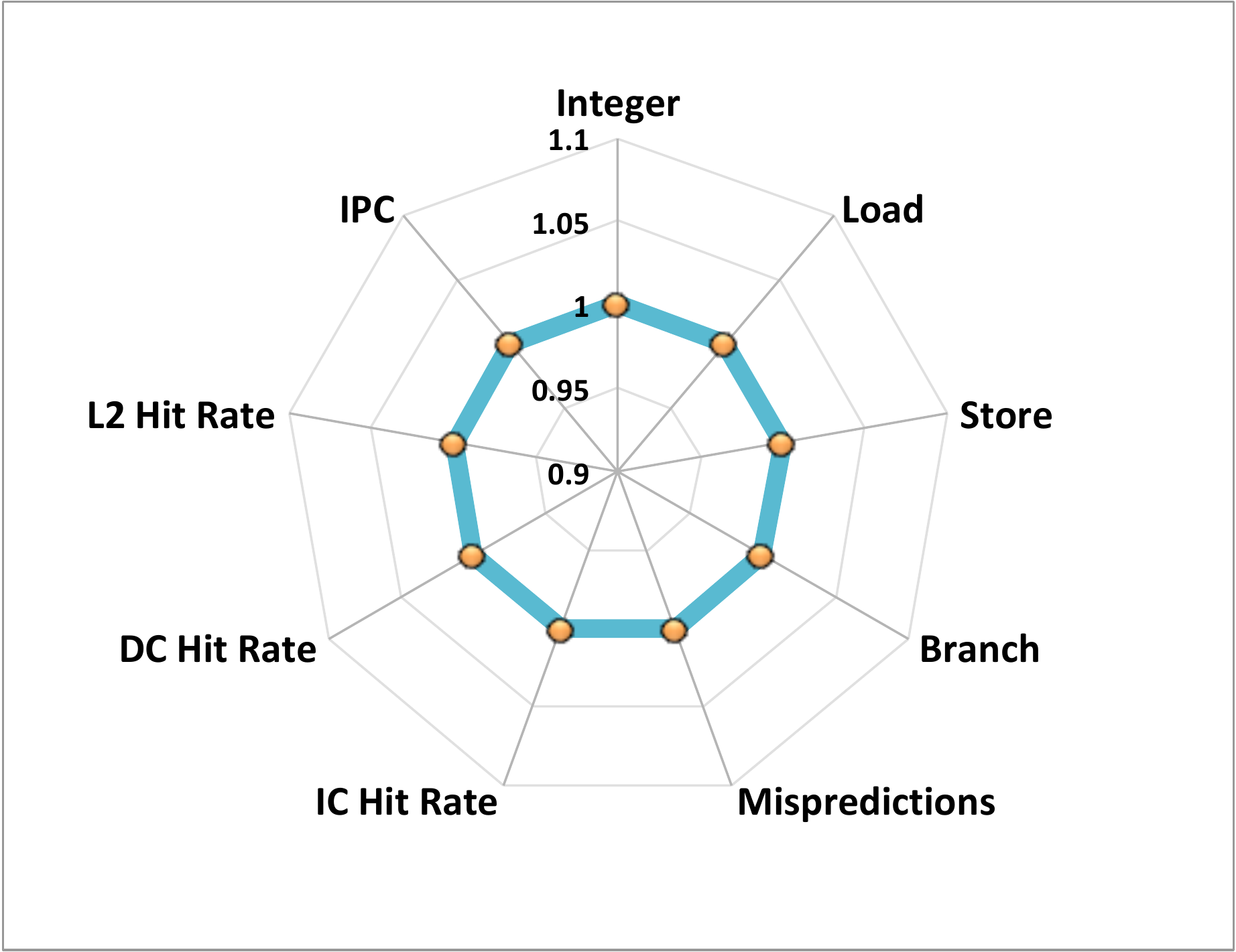}}
    \end{subfigure}
    \begin{subfigure}{0.24\textwidth}
    \fbox{\includegraphics[width=\textwidth,trim={1.25cm 2cm 2.5cm 1.25cm},clip]{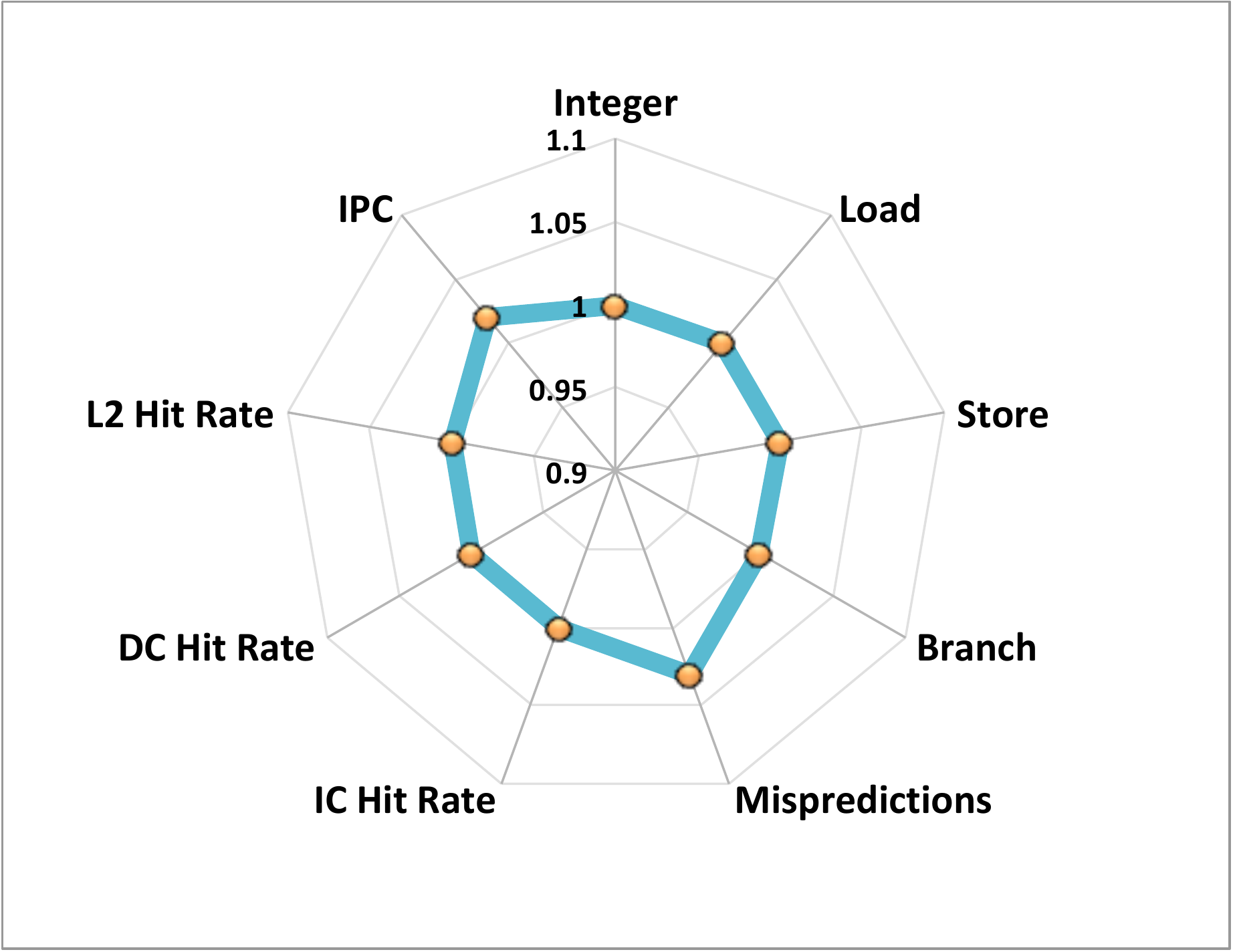}}
    \end{subfigure}
    \begin{subfigure}{0.24\textwidth}
    \fbox{\includegraphics[width=\textwidth,trim={1.25cm 2cm 2.5cm 1.25cm},clip]{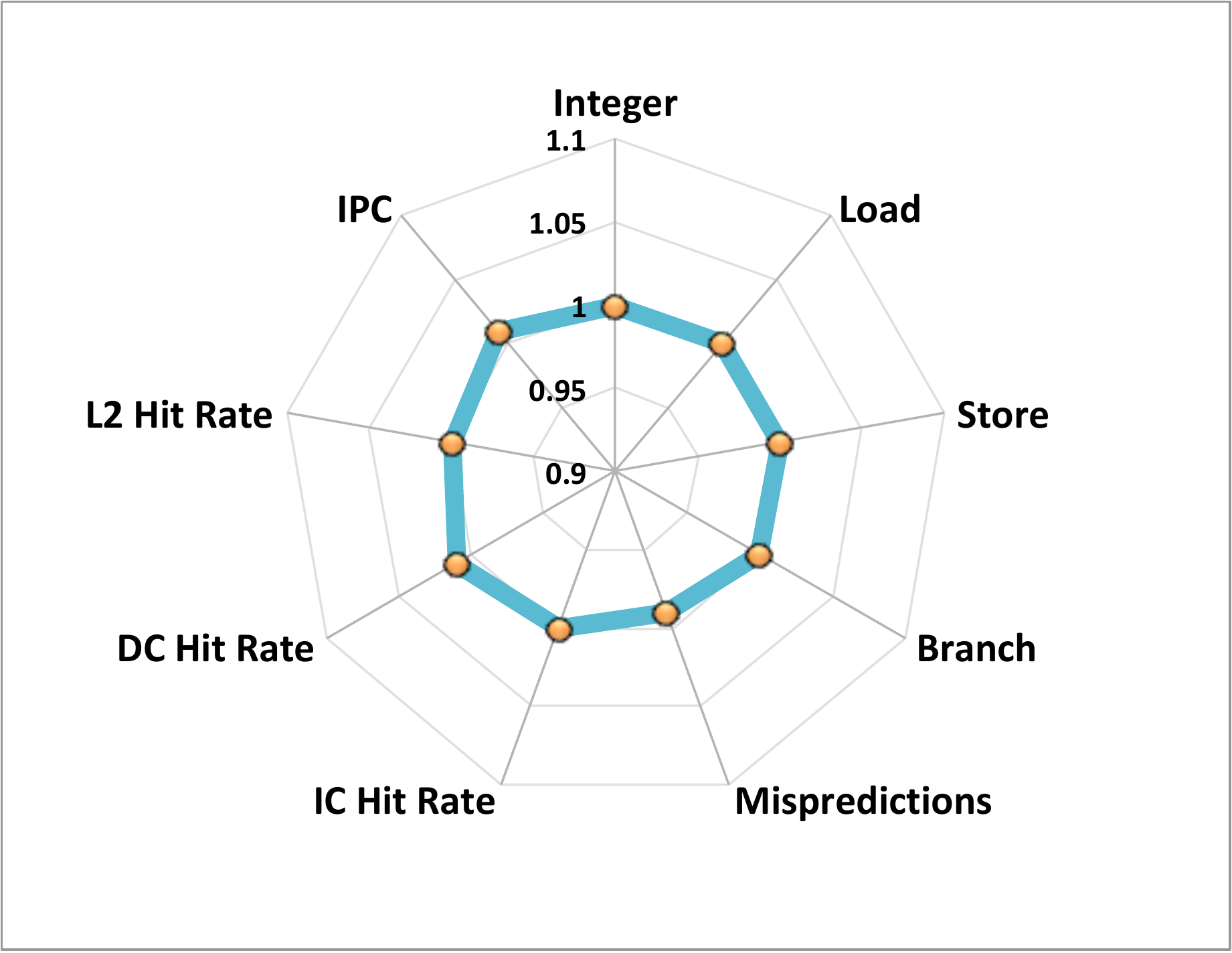}}
    \end{subfigure}
  \caption{Workload Cloning targeting a "large" core, with Gradient Descent. Top Left to Right Bottom: (a) astar [10 epochs], (b) bzip2 [5], (c) gcc [19]. (d) hmmer [52], (e) libquantm [45], (f) mcf [21], (g) sjeng [15], (h) xalancbmk [26] }
  \label{WC_GD_BIG}
\end{figure*}

\begin{figure*}[t]
  \centering
    \begin{subfigure}{0.24\textwidth}
    \fbox{\includegraphics[width=\textwidth,trim={1.25cm 2cm 2.5cm 1.25cm},clip]{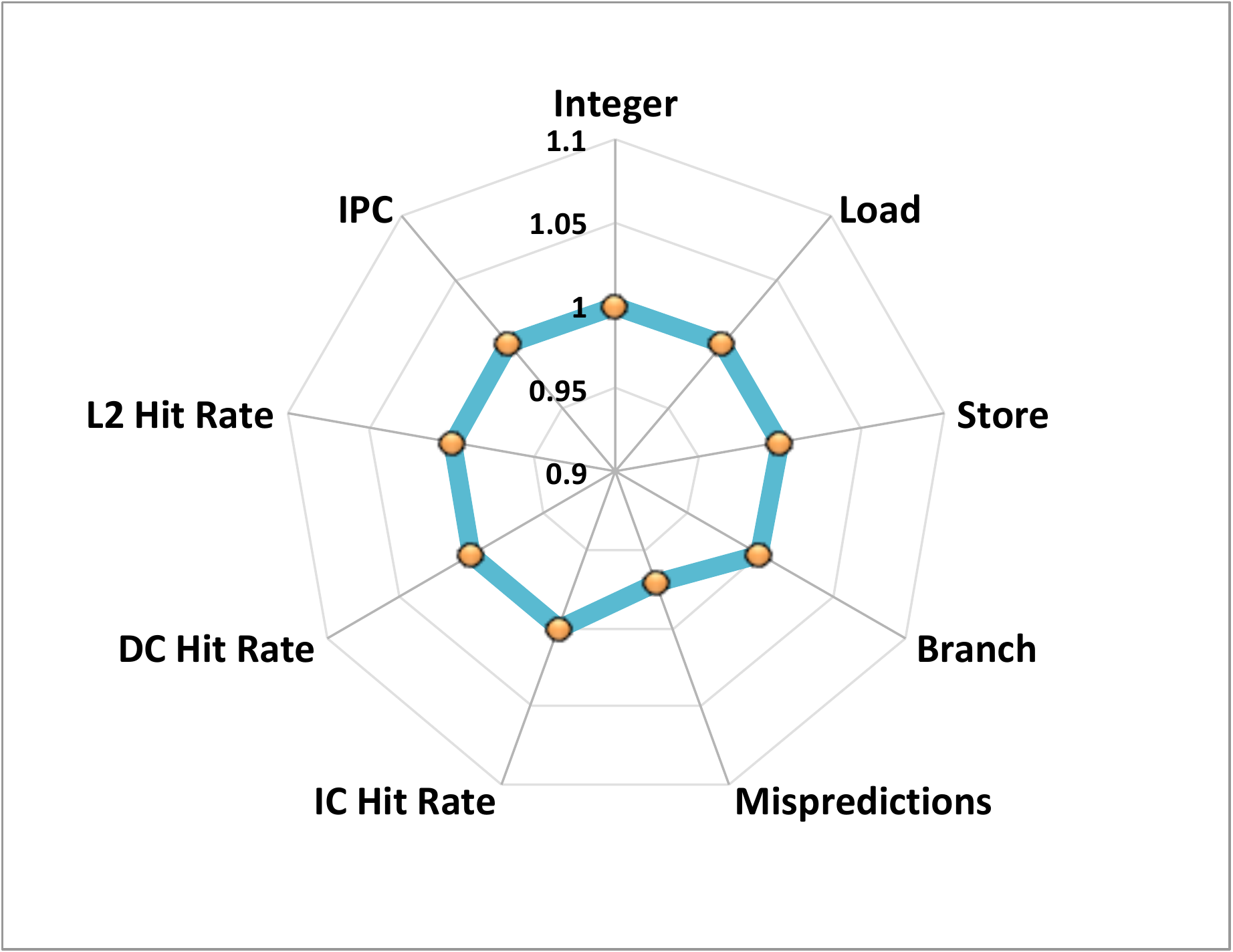}}
    \end{subfigure}
    \begin{subfigure}{0.24\textwidth}
    \fbox{\includegraphics[width=\textwidth,trim={1.25cm 2cm 2.5cm 1.25cm},clip]{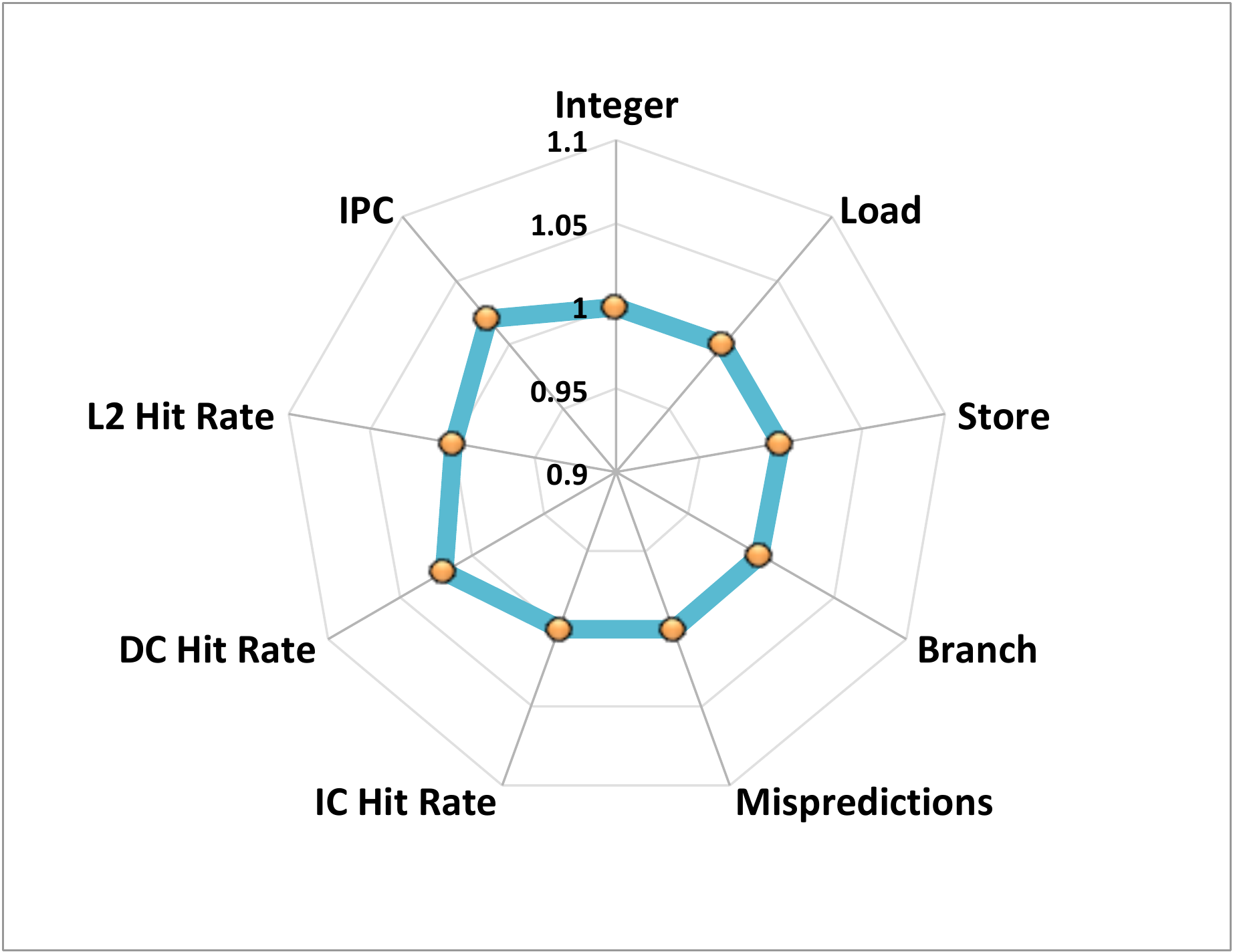}}
    \end{subfigure}
    \begin{subfigure}{0.24\textwidth}
    \fbox{\includegraphics[width=\textwidth,trim={1.25cm 2cm 2.5cm 1.25cm},clip]{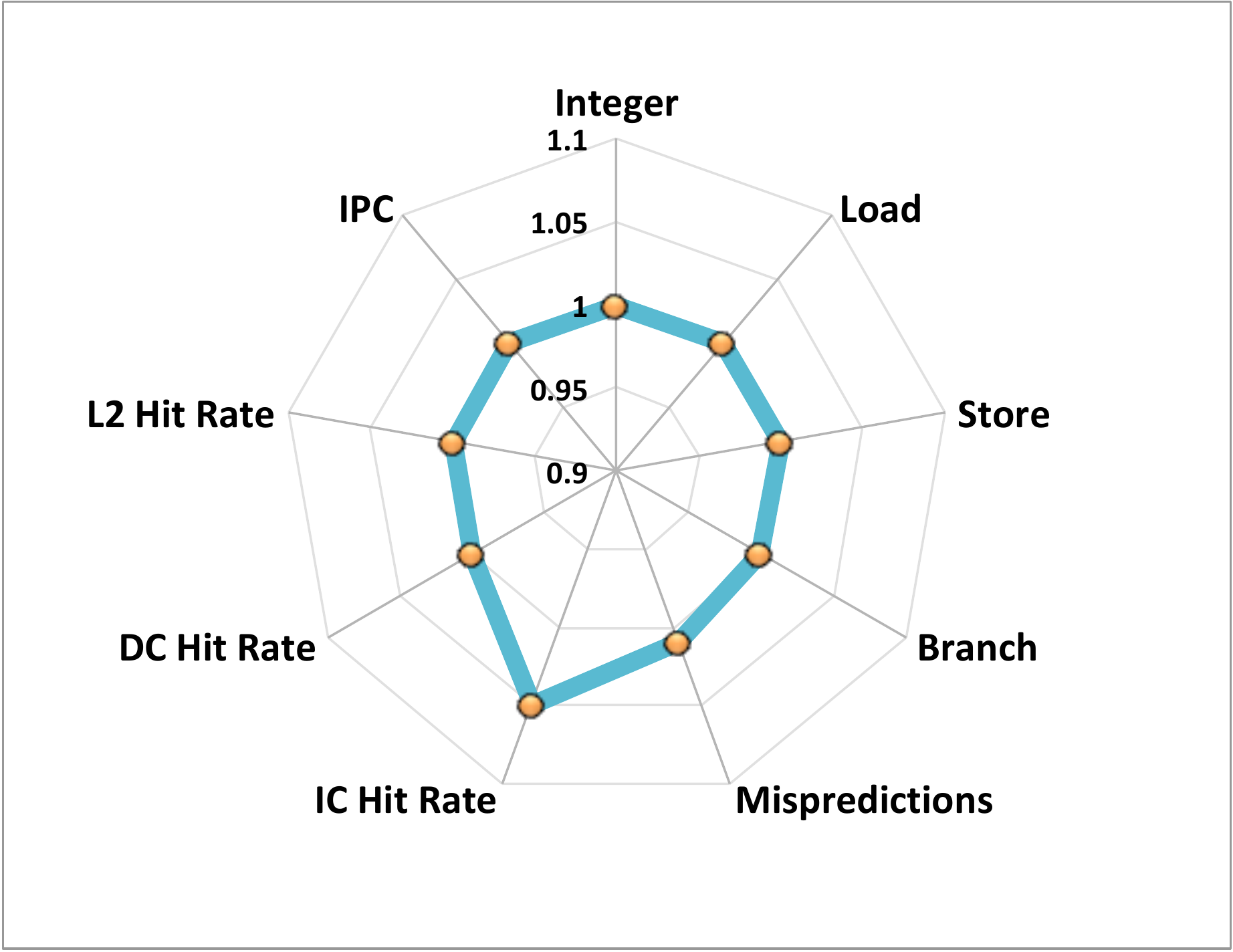}}
    \end{subfigure}
    \begin{subfigure}{0.24\textwidth}
    \fbox{\includegraphics[width=\textwidth,trim={1.25cm 2cm 2.5cm 1.25cm},clip]{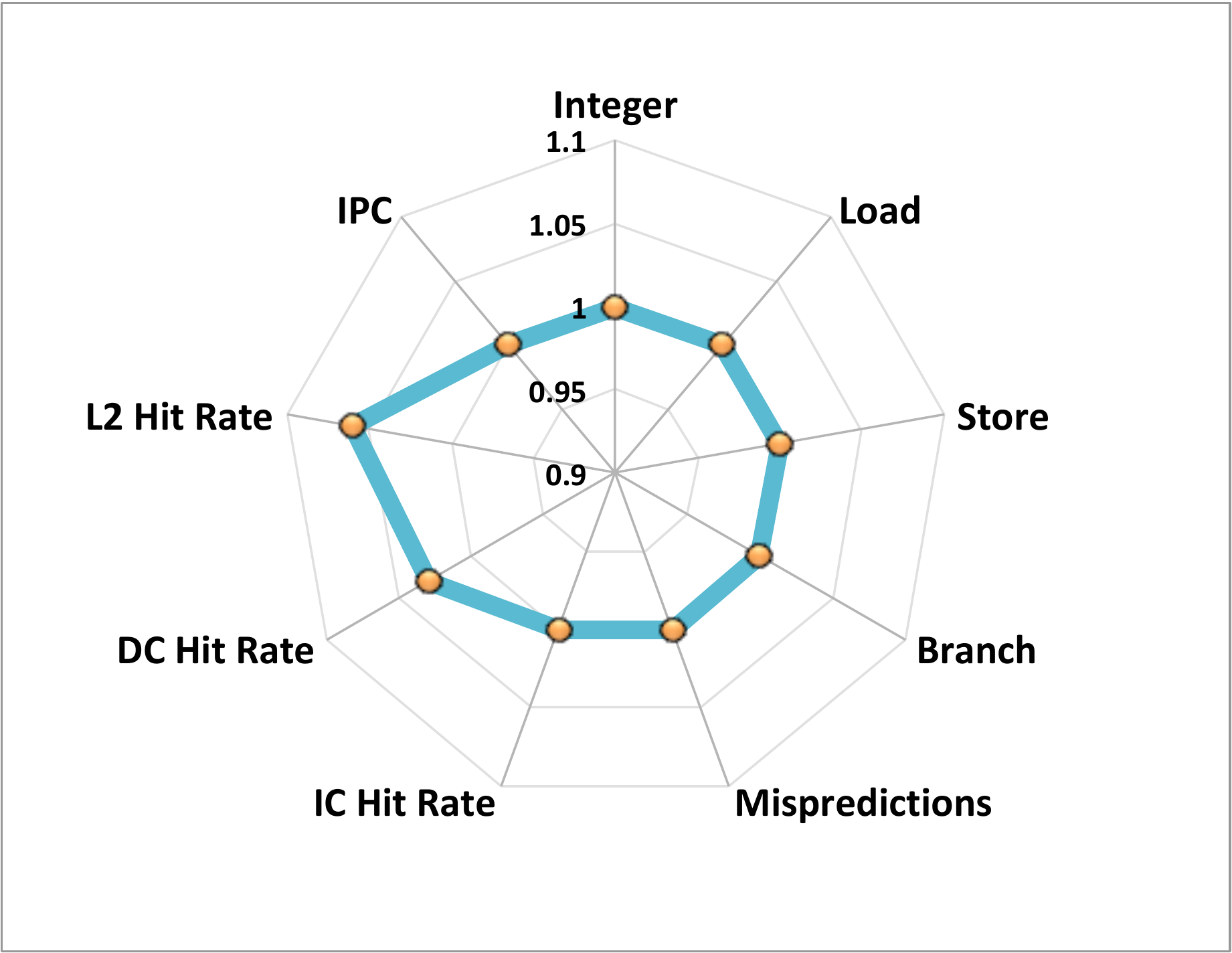}}
    \end{subfigure}
    \begin{subfigure}{0.24\textwidth}
    \fbox{\includegraphics[width=\textwidth,trim={1.25cm 2cm 2.5cm 1.25cm},clip]{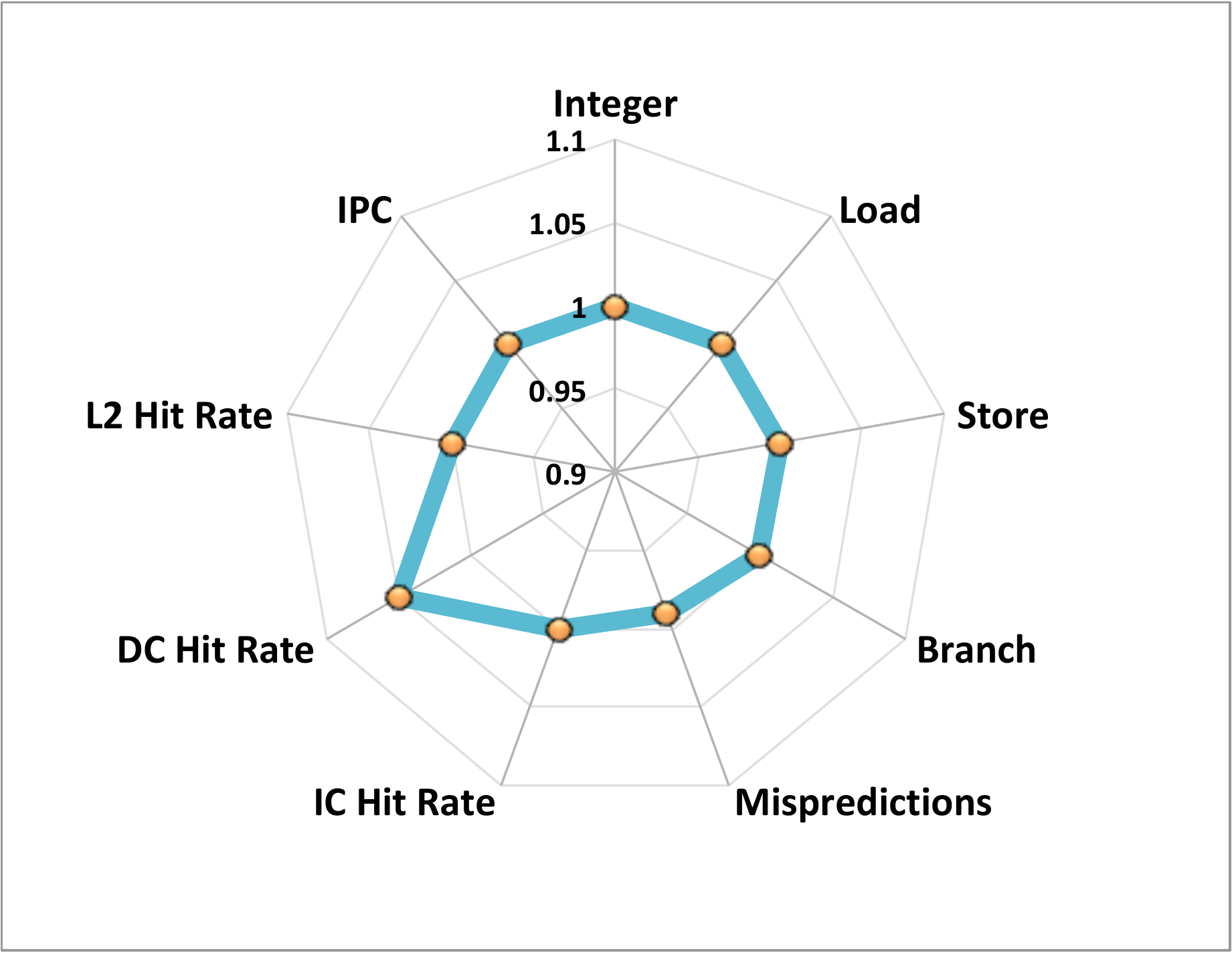}}
    \end{subfigure}
    \begin{subfigure}{0.24\textwidth}
    \fbox{\includegraphics[width=\textwidth,trim={1.25cm 2cm 2.5cm 1.25cm},clip]{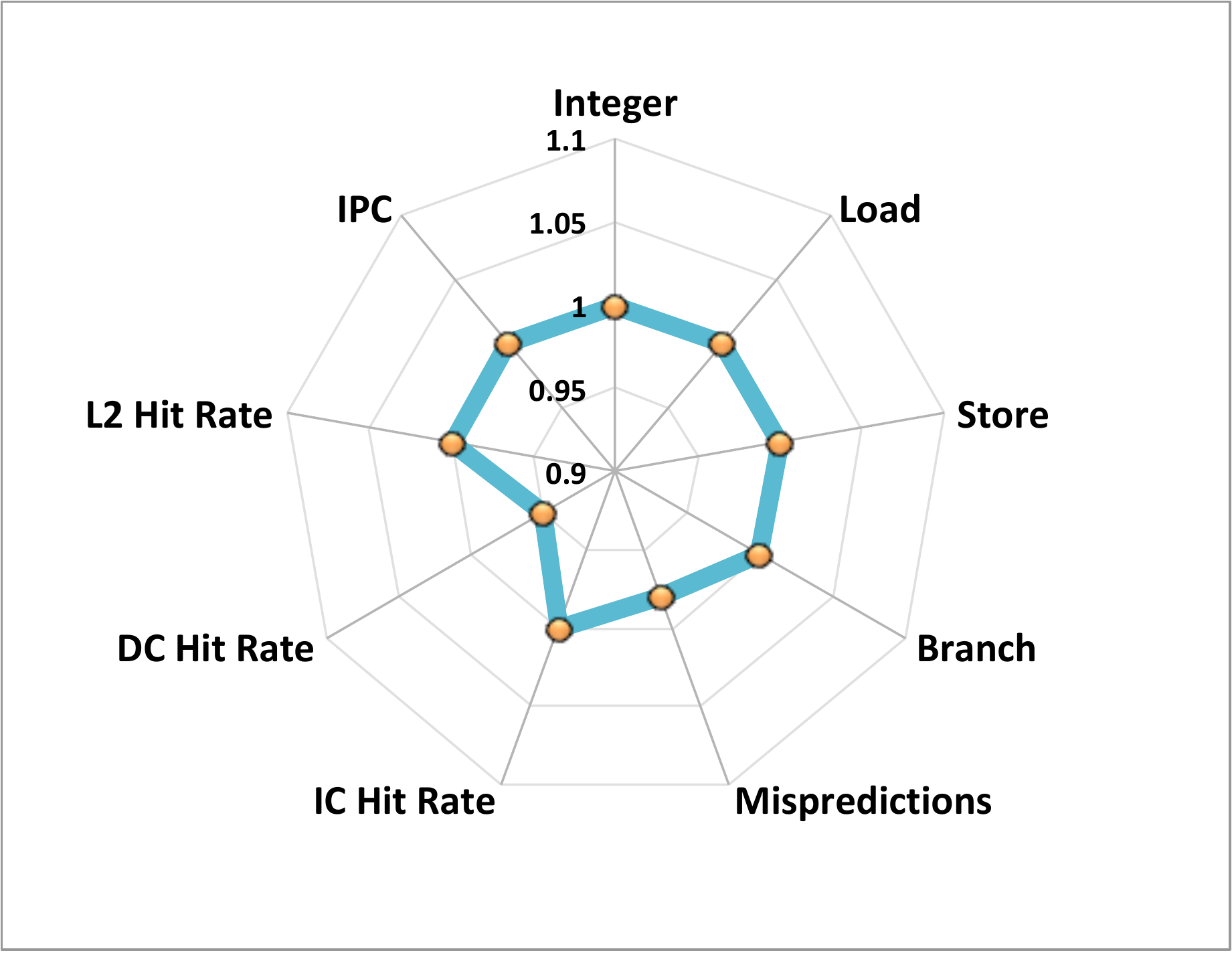}}
    \end{subfigure}
    \begin{subfigure}{0.24\textwidth}
    \fbox{\includegraphics[width=\textwidth,trim={1.25cm 2cm 2.5cm 1.25cm},clip]{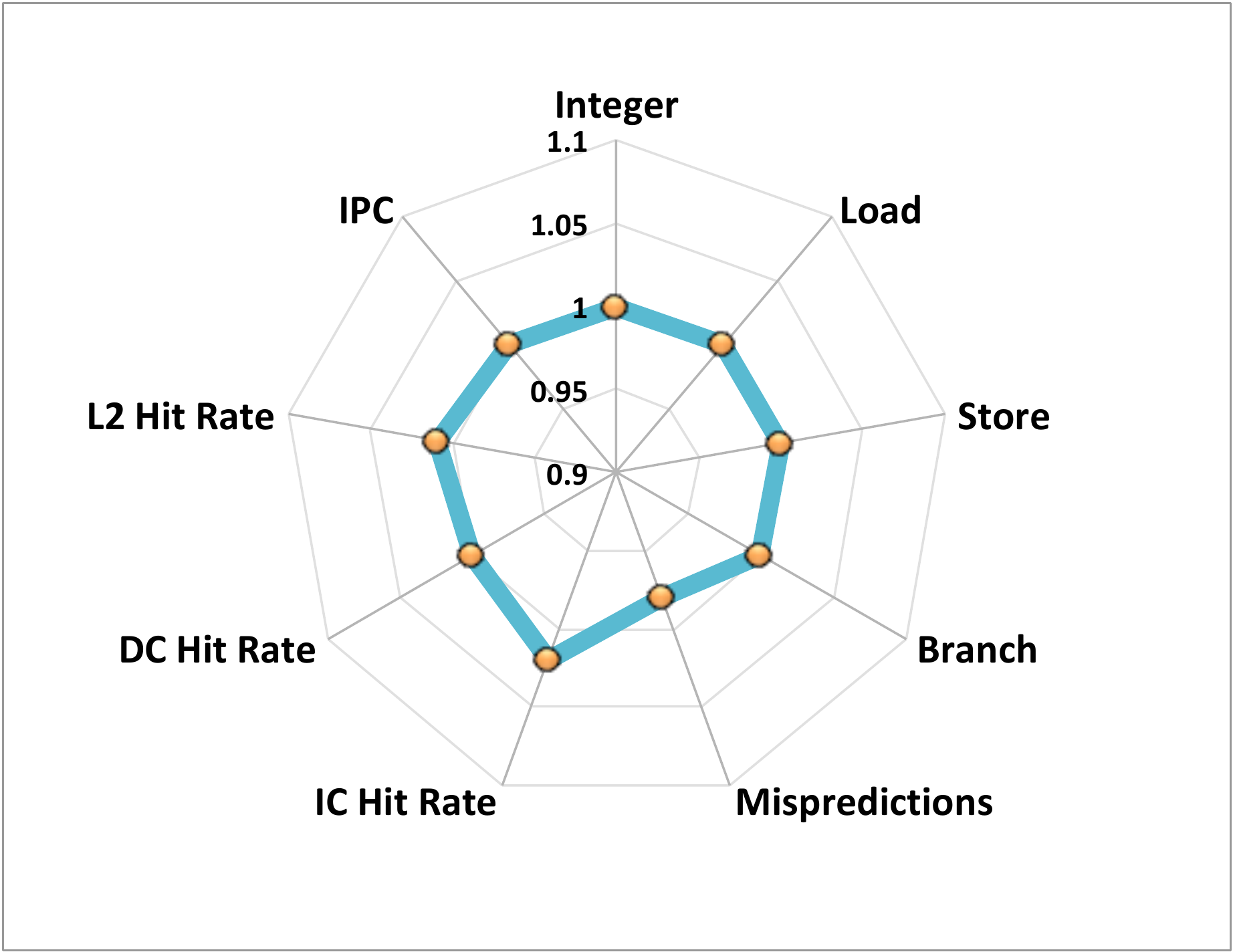}}
    \end{subfigure}
    \begin{subfigure}{0.24\textwidth}
    \fbox{\includegraphics[width=\textwidth,trim={1.25cm 2cm 2.5cm 1.25cm},clip]{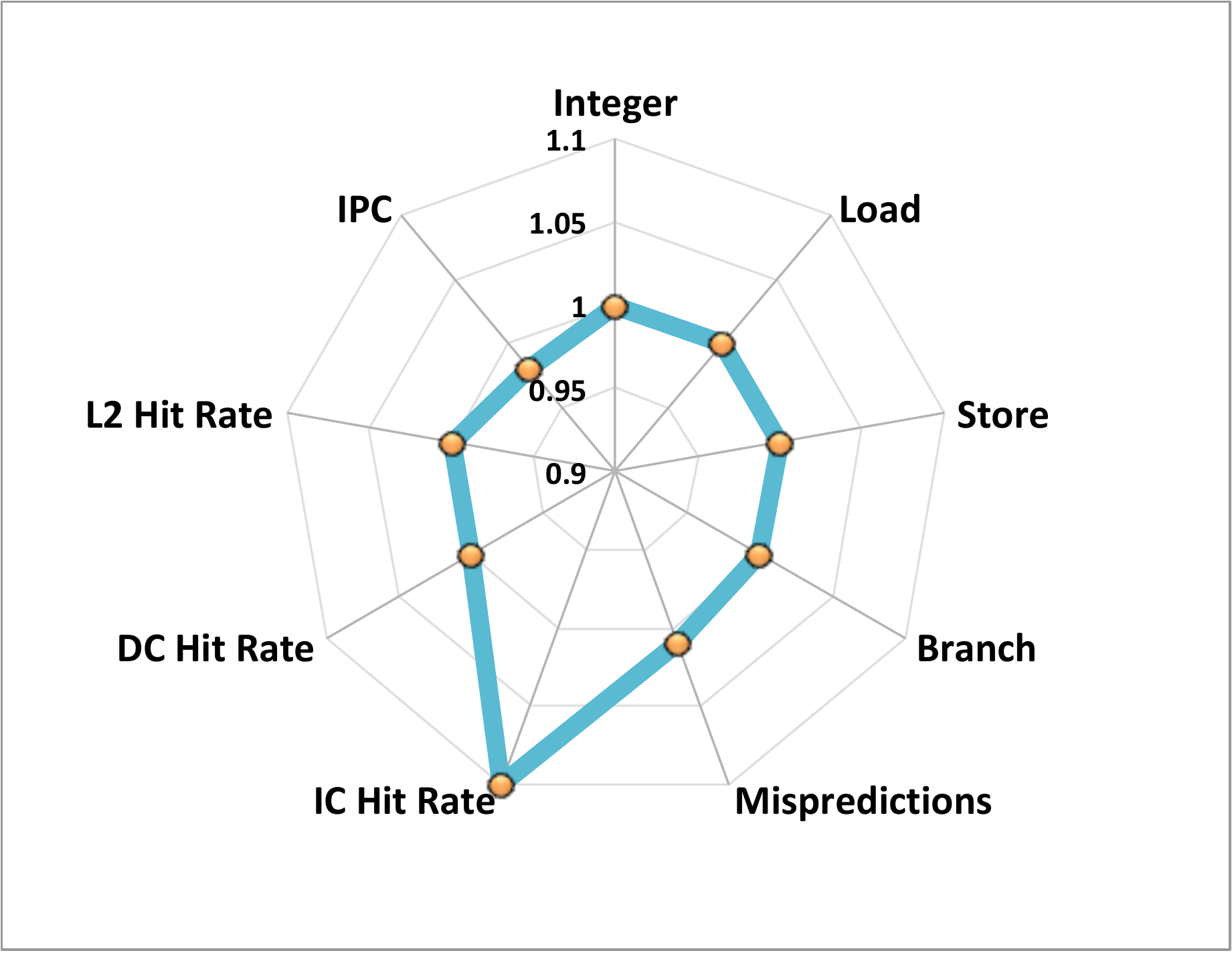}}
    \end{subfigure}
  \caption{Workload Cloning targeting a "small" core, with Gradient Descent. Top Left to Right Bottom: (a) astar [21 epochs], (b) bzip2 [5], (c) gcc [36]. (d) hmmer [40], (e) libquantm [50], (f) mcf [30], (g) sjeng [6], (h) xalancbmk [37] }
  \label{WC_GD_LITTLE}
\end{figure*}

\subsection{Experimental Setup}
\subsubsection{Workloads}
To evaluate the cloning use case, we choose 8 benchmarks from the SPEC INT CPU2006~\cite{spec} suite and generate clones on simpoints~\cite{simpoint} of 100 million instructions. 
The generated test cases (for both use cases) are made up of roughly 500 static instructions in an endless loop and run for a total of 10 million dynamic instructions.

\subsubsection{Evaluation Framework}
We target the Gem5~\cite{gem5} architectural performance simulator and the McPAT~\cite{mcpat} power estimation framework.
While performance numbers and module level statistics can be evaluated from Gem5 alone, power estimation requires the transfer of execution statistics from Gem5 to McPAT, based on which dynamic power is estimated. 

\subsubsection{Target Microarchitectures}
We target the RISC-V ISA. We model two cores --\emph{Large} and \emph{Small}-- to evaluate the performance of MicroGrad on different corners of the architecture design space. 
The details of each core are listed in Table \ref{Core}.
For the power template, we use the default McPAT configurations commensurate with these core sizes.

\subsubsection{Metrics / Accuracy}
For Workload Cloning, we focus on: i) Integer, Branch, Load, Store instructions, ii) L1D, L1I, L2 cache hit rates, iii) Branch misprediction rate and iv) IPC.
For Stress Testing, we focus separately on IPC and Dynamic Power.
The Loss function utilized by the tuning algorithm calculates log loss over the metrics of interest specified above.
Where applicable, we target an accuracy of 99\% across the metrics.

\begin{figure*}[t]
  \centering
    \begin{subfigure}{0.24\textwidth}
    \fbox{\includegraphics[width=\textwidth,trim={1.25cm 2cm 2.5cm 1.25cm},clip]{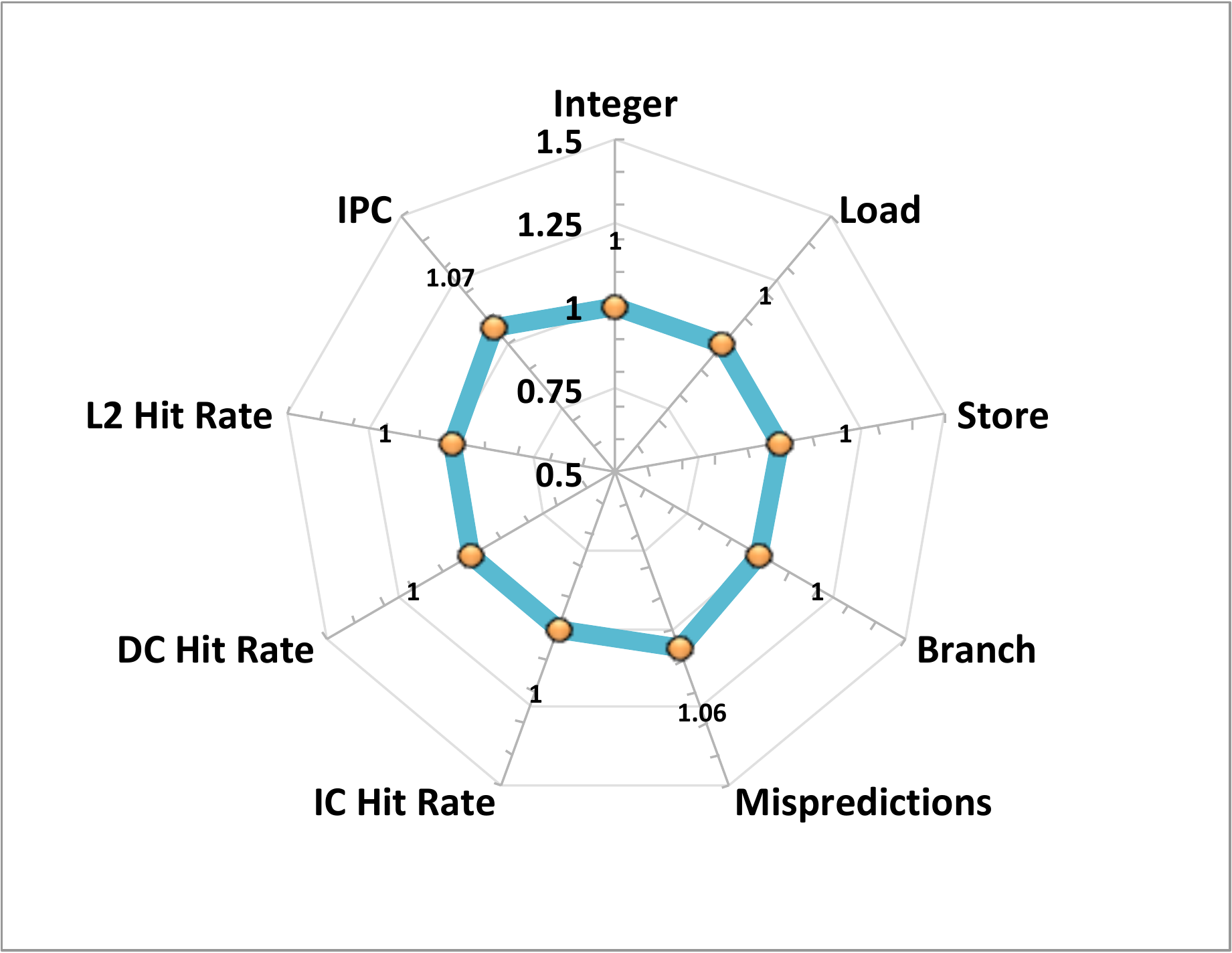}}
    \end{subfigure}
    \begin{subfigure}{0.24\textwidth}
    \fbox{\includegraphics[width=\textwidth,trim={1.25cm 2cm 2.5cm 1.25cm},clip]{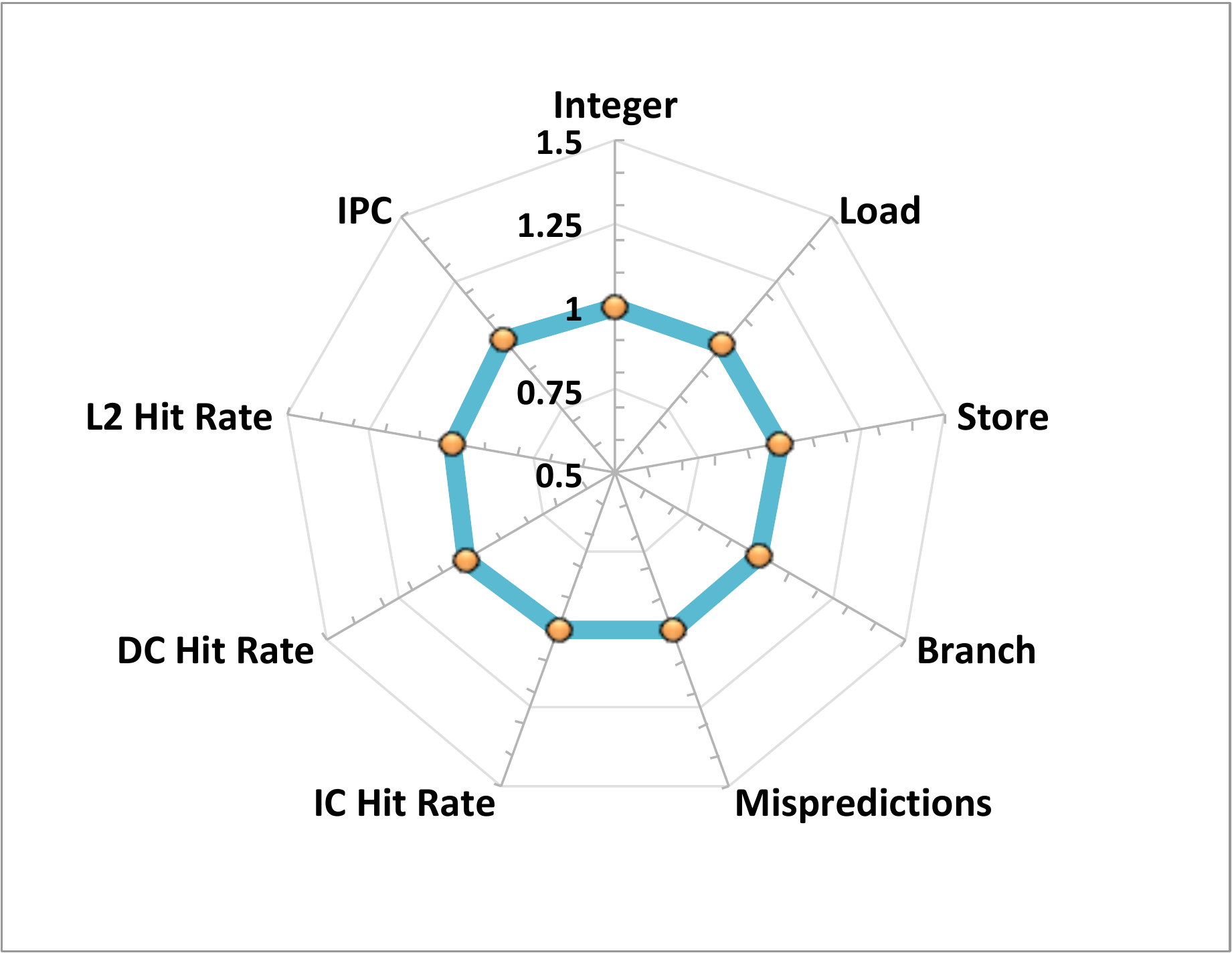}}
    \end{subfigure}
    \begin{subfigure}{0.24\textwidth}
    \fbox{\includegraphics[width=\textwidth,trim={1.25cm 2cm 2.5cm 1.25cm},clip]{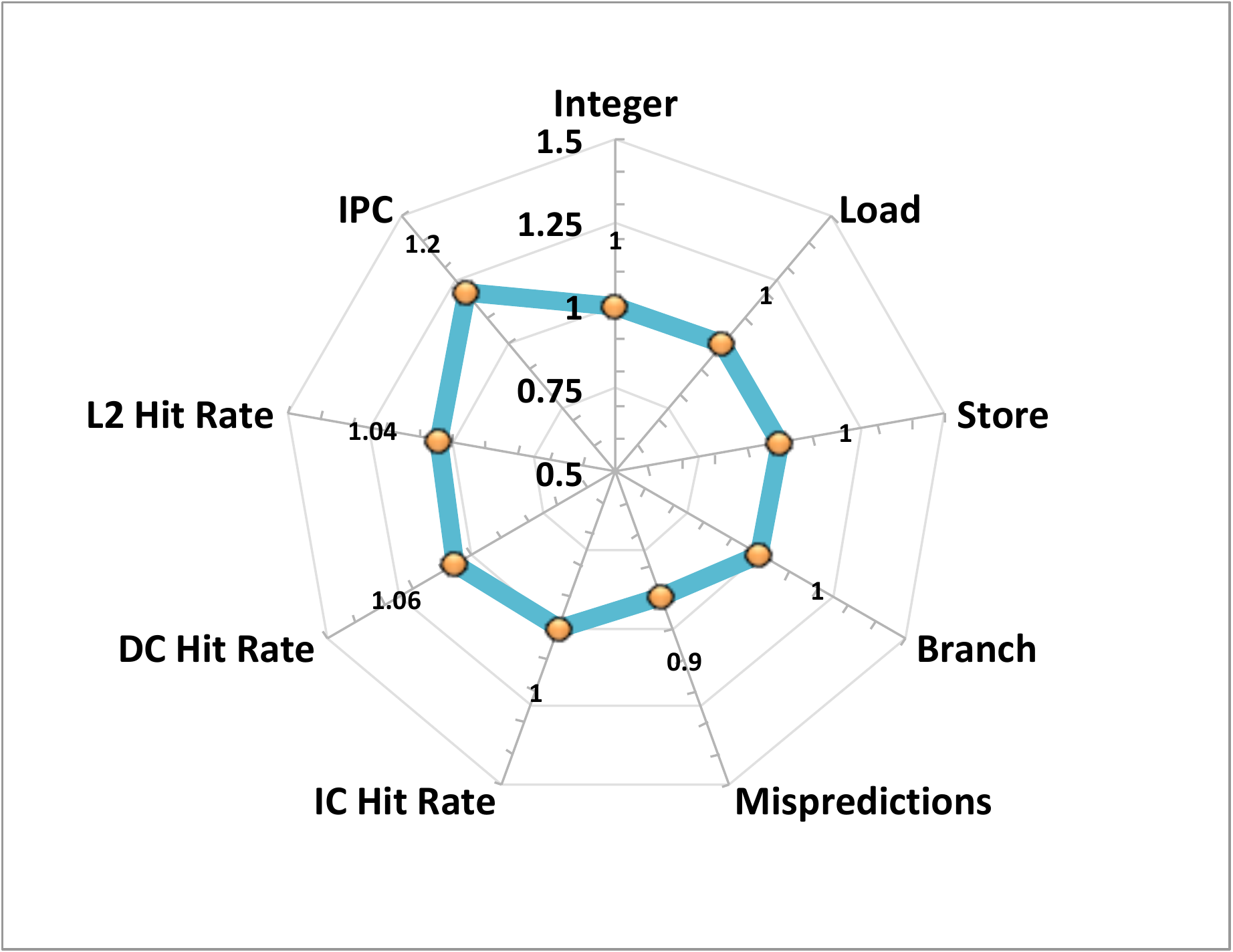}}
    \end{subfigure}
    \begin{subfigure}{0.24\textwidth}
    \fbox{\includegraphics[width=\textwidth,trim={1.25cm 2cm 2.5cm 1.25cm},clip]{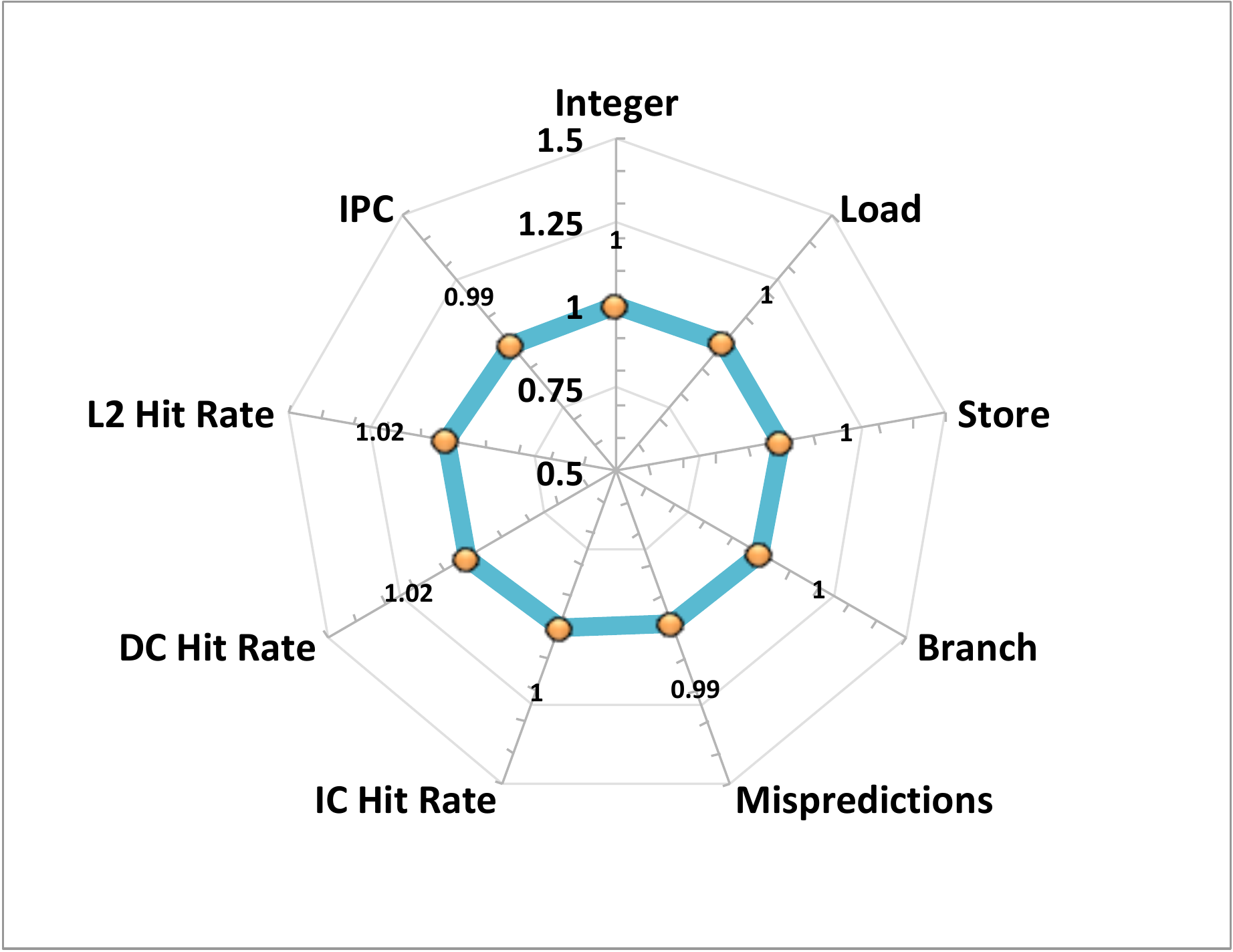}}
    \end{subfigure}
    \begin{subfigure}{0.24\textwidth}
    \fbox{\includegraphics[width=\textwidth,trim={1.25cm 2cm 2.5cm 1.25cm},clip]{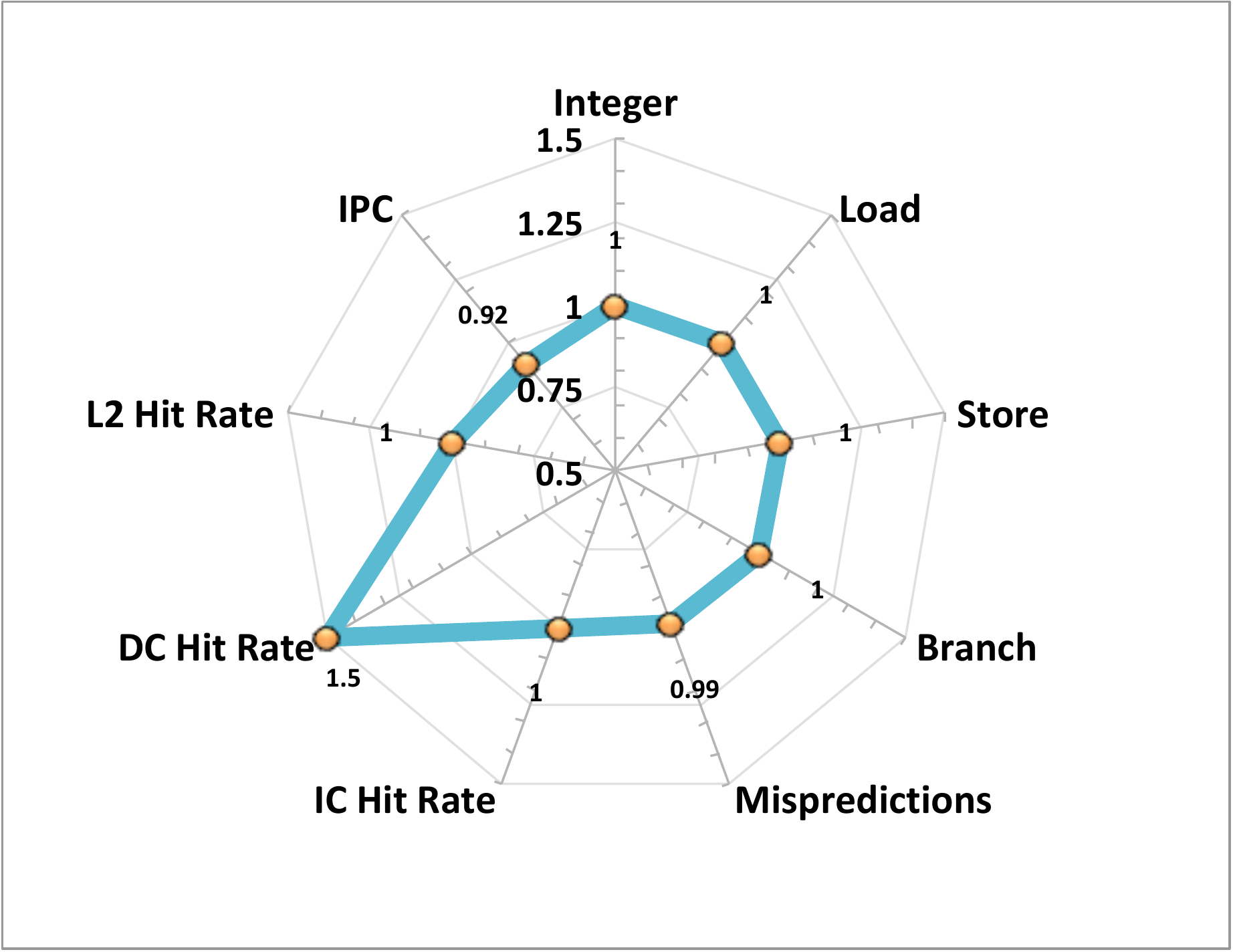}}
    \end{subfigure}
    \begin{subfigure}{0.24\textwidth}
    \fbox{\includegraphics[width=\textwidth,trim={1.25cm 2cm 2.5cm 1.25cm},clip]{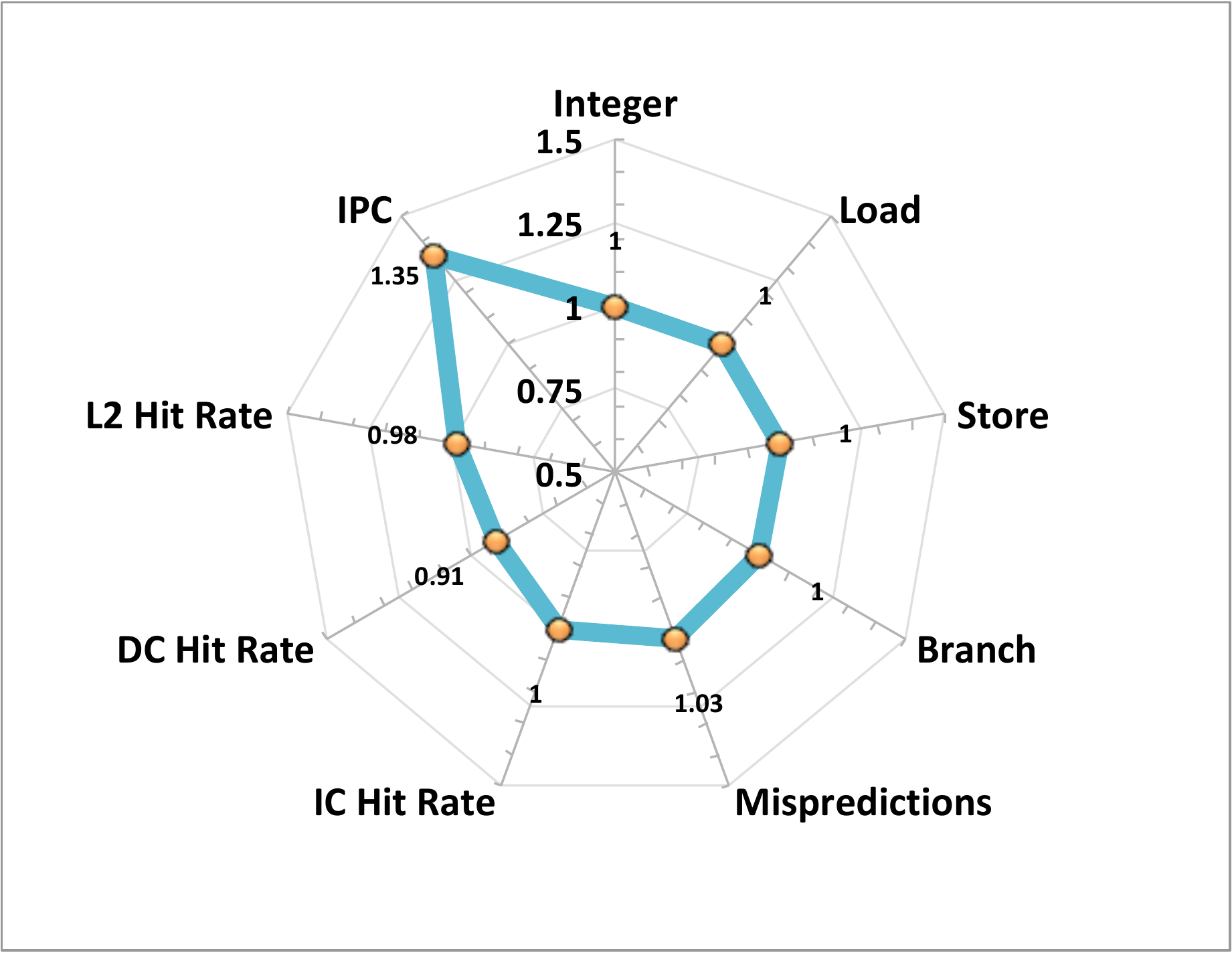}}
    \end{subfigure}
    \begin{subfigure}{0.24\textwidth}
    \fbox{\includegraphics[width=\textwidth,trim={1.25cm 2cm 2.5cm 1.25cm},clip]{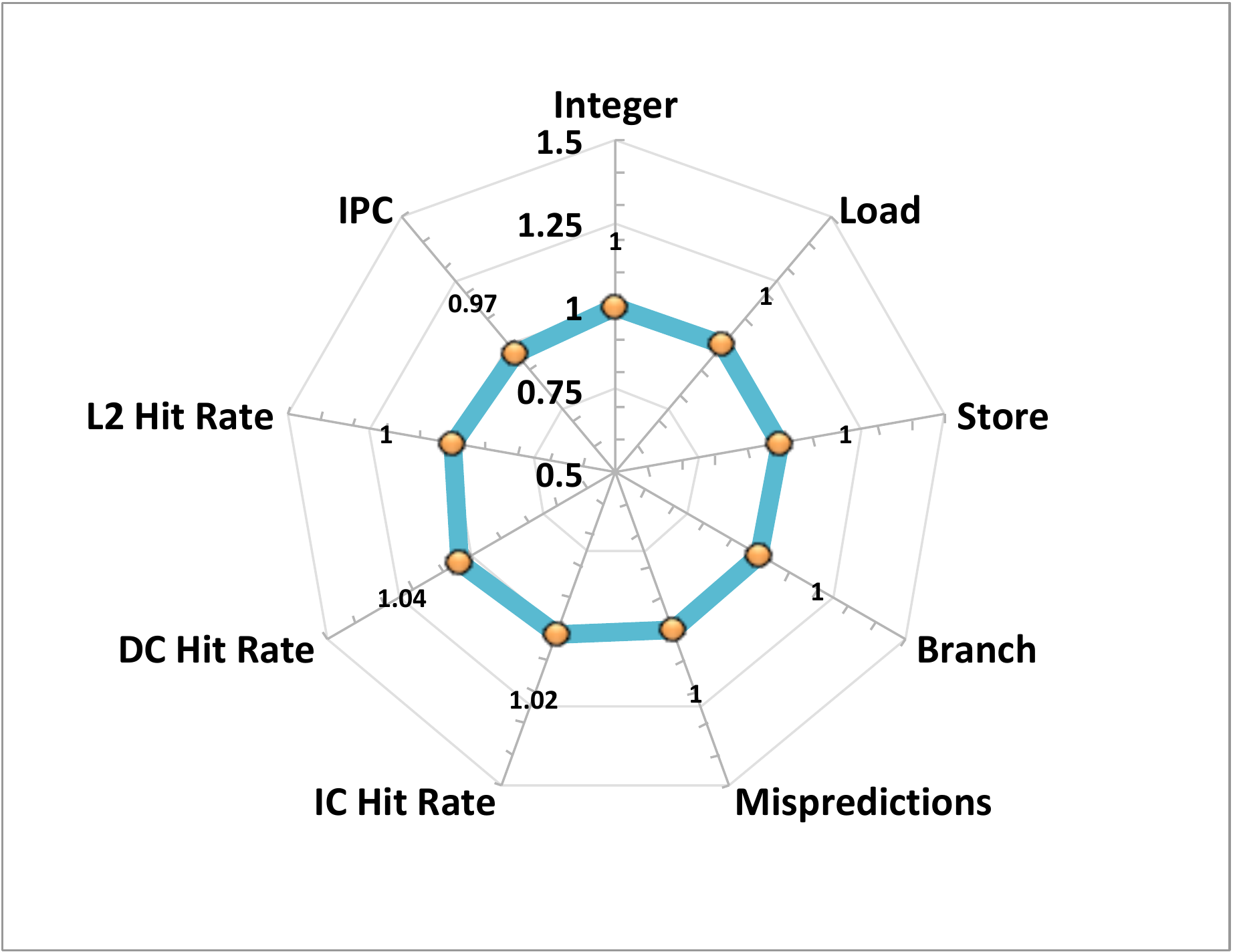}}
    \end{subfigure}
    \begin{subfigure}{0.24\textwidth}
    \fbox{\includegraphics[width=\textwidth,trim={1.25cm 2cm 2.5cm 1.25cm},clip]{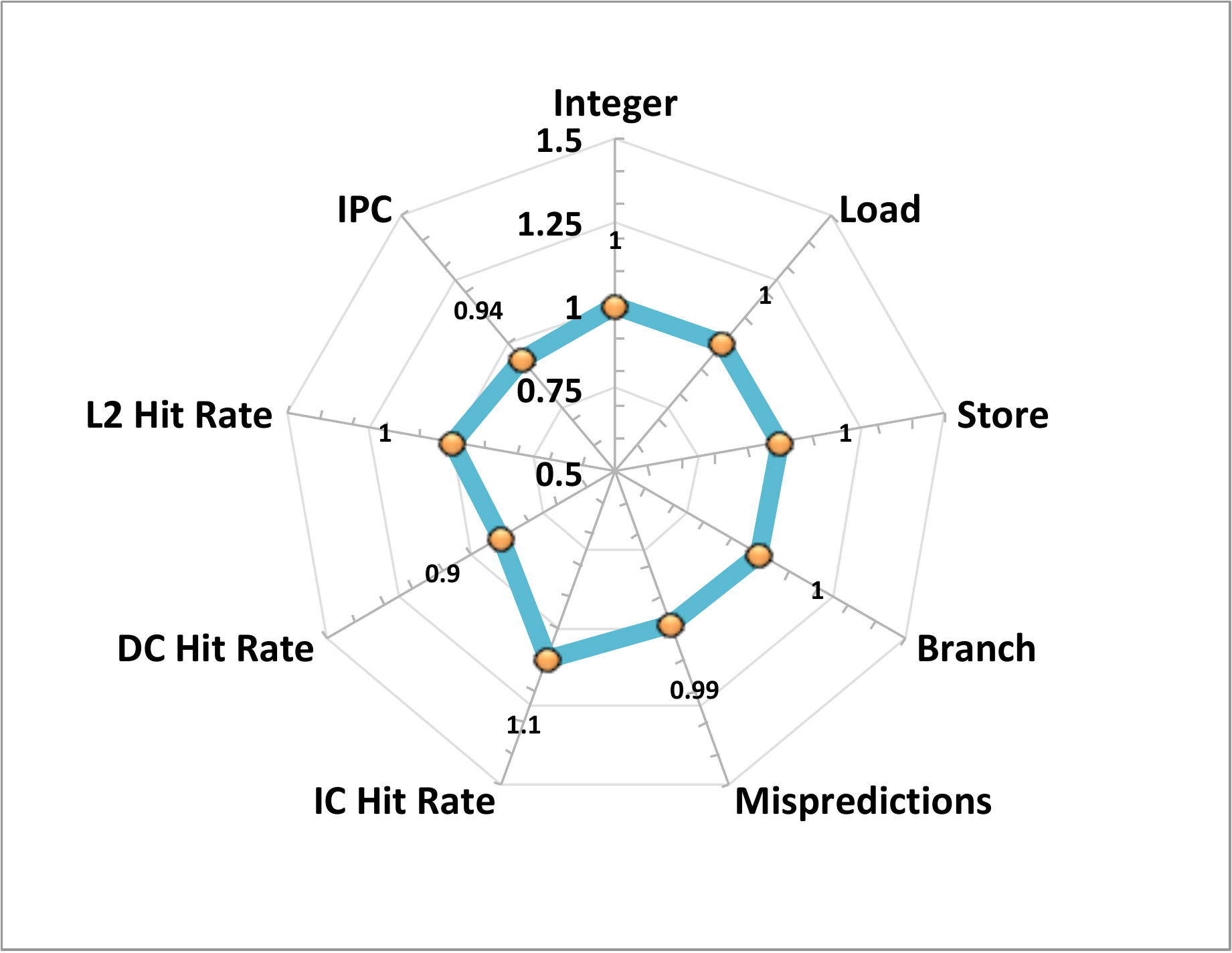}}
    \end{subfigure}
  \caption{Workload Cloning targeting a "large" core, with Genetic Algorithm. Top Left to Right Bottom: (a) astar [10 epochs], (b) bzip2 [5], (c) gcc [19]. (d) hmmer [52], (e) libquantm [45], (f) mcf [21], (g) sjeng [15], (h) xalancbmk [26] }
  \label{WC_GA_BIG}
\end{figure*}

\subsection{Workload Cloning}
In Fig.\ref{WC_GD_BIG} and Fig.\ref{WC_GD_LITTLE} we showcase the efficiency of MicroGrad towards Workload Cloning.
Fig.\ref{WC_GD_BIG} shows the workload clones generated across the 8 benchmarks on a \emph{Large} core while Fig.\ref{WC_GD_LITTLE} shows the same on a \emph{Small} core.
In the figures, the circumferential axis represents different metrics - instructions distributions, mispredictions, cache miss rates and IPC.
The radial axis represents the accuracy of the clone's metric compared to the original benchmark (1 indicates complete accuracy).


For the \emph{Large} core, over the eight benchmarks, the accuracy across all metrics is close to 1 (average error is less than 1\%).
Worse case scenario is seen in libquantum wherein there is close to a 5\% error in the branch misprediction rate and the data cache (DC) hit rate.

In the case of the \emph{Small} core, results are similar (average error is less than 2\%).
The accuracy is marginally less compared to the \emph{Large} core due to the higher metric sensitivity in a core of smaller size.
This is due to program characteristics having a larger impact on the execution flow, since the core is not over provisioned with resources.
The worse case error is close to 10\% in the case of xalancbmk's IC hit rate.
We note that there is potential for more knobs to be implemented in MicroGrad that can control IC Hit Rates with higher accuracy, which we seek to implement in the future.

The captions of both figures indicate the number of epochs required to create the workload clones.
Epochs vary from only 5, to a maximum of 52, clearly highlighting that MicroGrad's high accuracy is achievable in very few tuning epochs.

The accuracy and fast tuning capability of MicroGrad is heavily influenced by the Gradient Descent tuning algorithm.
To showcase this, we compare against a Genetic Algorithm based approach in Fig.\ref{WC_GA_BIG} for the Big core.
The GA parameters are taken from prior work and were shown in Table \ref{GA-param}.
For this analysis, we allow the GA based approach to run for the same number of tuning epochs as the GD based approach. 
The figure shows that the accuracy achieved by GA is considerably lower than the GD approach (note that the ratios on the radial axes are far greater).
The average error in comparison to the original benchmarks is roughly 30\%, with worst case errors of more than 50\%

It should also be noted that allowing the same number of epochs is favorable to GA.
As discussed earlier, the GA tuning epoch (with Table \ref{GA-param} parameters) performs roughly 2.5 times the work of the GD based approach: 50 evaluations per epoch (population size) in GA vs 20 evaluations per epoch (2 x knobs) in GD.
Depending on the implementation, this can manifest as higher execution time, more compute resources needed or both.

Also significant to note is that the GA based tuning algorithm fits seamlessly into the MicroGrad framework. This is thanks to the modular implementation of MicroGrad which allows for flexible development on multiple fronts, including the research on use case specific tuning algorithms.

\begin{figure}[thp]
\includegraphics[width=\columnwidth,trim={0cm 3.5cm 0cm 0cm},clip]{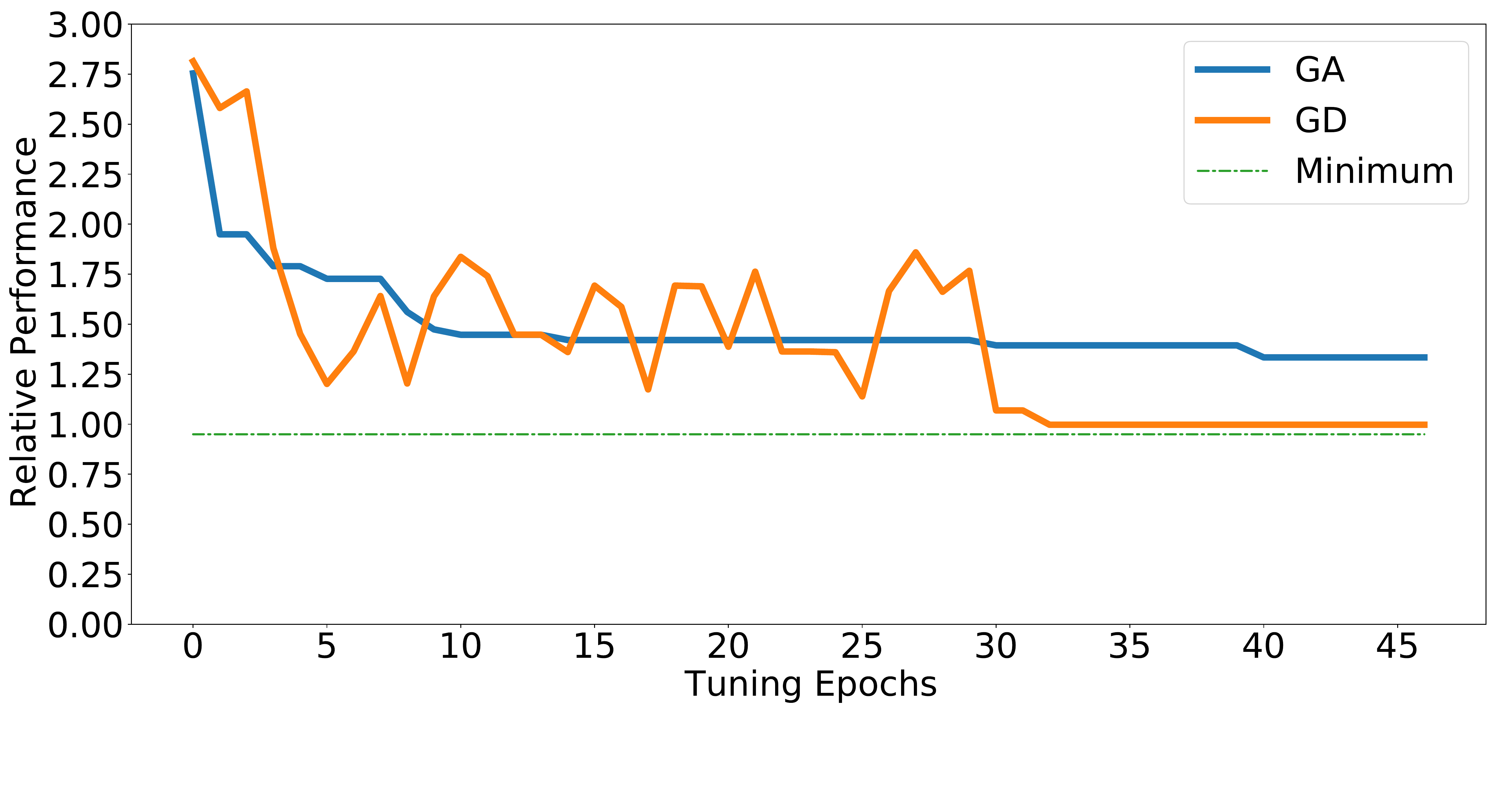}
\centering
\caption{Performance virus: GD vs GA}
\label{ST_Perf_GDGA}
\end{figure}

\subsection{Stress Testing}
Next, we discuss MicroGrad's proficiency in stress testing.
Fig.\ref{ST_Perf_GDGA} shows a compute-focused performance stress test scenario which seeks to achieve the worst case performance on the \emph{Large} core.
This testing scenario is focused only on tuning the instruction fractions and not on other metrics like miss rates and mispredictions.
The green line shows the optimal worst case performance as estimated by a brute-force search exploring the entire workload space.
The Gradient Descent mechanism (shown in orange), is able to converge to the worst case in under 30 epochs. 
In comparison, a GA based tuning approach (green) is about 25\% off from the optimal worse case performance in 1.5 times the number of epochs.

\begin{figure}[thp]
\includegraphics[width=\columnwidth,trim={0cm 5cm 0cm 0cm},clip]{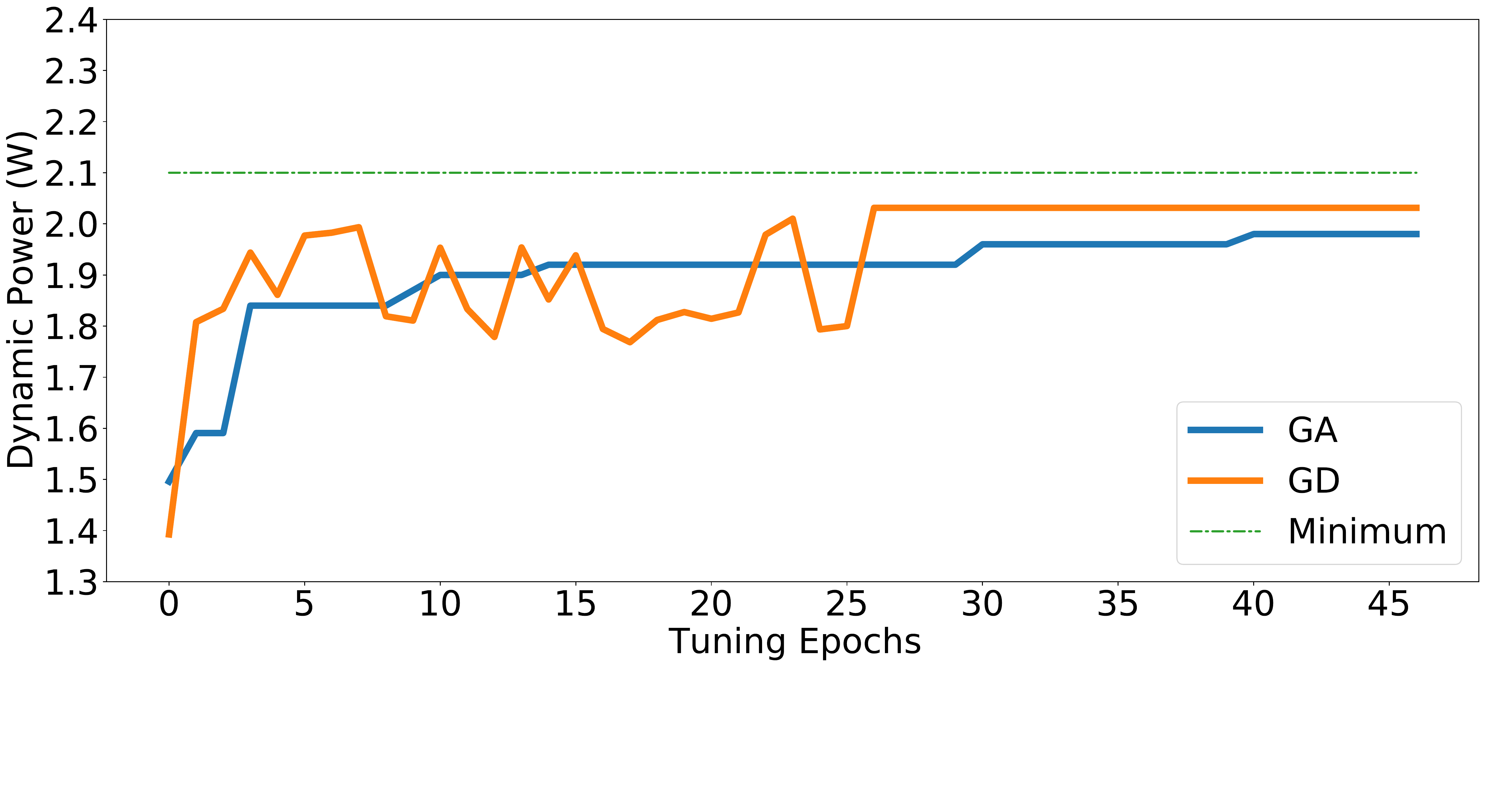}
\centering
\caption{Power virus: GD vs GA}
\label{ST_Power_GD}
\end{figure}

Next, in Fig.\ref{ST_Power_GD} we show a compute-focused stress test scenario targeting worst case dynamic power.
Again, the green line shows the highest dynamic power achieved through brute-force search across the workload space - roughly 2.1 W.
The GD approach is able to achieve 2.01 W (95\% accuracy) in only 25 tuning epochs.
In comparison, the GA approach is able to achieve power that is similar to GD, but requires roughly 2x the number of epochs.

Further, in Table \ref{PV-ID} we show the distribution of instructions in the GD generated power virus - which shows similarity to the result of the brute-force search.
More than 50\% of the instructions are memory focused and over 20\% are floating point operations.
On the other hand, the integer operations are only 6\% of the total. 
The high fractions for memory and FP ops are intuitive considering that these operations perform more complex microarchitectural activity compared to integer operations.
Further (not shown), the register dependency distance chosen by this stress test was at its maximum limit, meaning that ILP was pushed to the maximum extent allowed.
This is also intuitive - more the microarchitectural activity, higher the power consumed.

Overall, these results indicate that gradient based tuning approach, in combination with an abstract workload model, can generate highly accurate stress tests on different use cases. 
In addition, the gradient decent tuning outperforms existing GA-based solutions in terms of time to a solution (epochs) and efficiency in resource utilization. 

\begin{table}[h]
\centering
\resizebox{0.85\columnwidth}{!}{%
\begin{tabular}{|l|l|l|l|l|}
\hline
\textbf{Integer} & \textbf{Float} & \textbf{Branch} & \textbf{Load} & \textbf{Store} \\ \hline
5.7\%            & 22.8\%         & 14.3\%          & 22.8\%        & 32.8\%         \\ \hline
\end{tabular}%
}
\caption{Power virus: Instruction Distribution}
\label{PV-ID}
\end{table}

\section{Conclusion}
In summary, we presented MicroGrad, an open-source centralized framework for workload cloning and stress testing.
Key novel features in MicroGrad are its gradient based tuning approach and its Microprobe back-end.
The framework is able to produce fast and accurate workload clones and stress tests.
These results are especially evident in comparison to prior techniques.

Beyond the specific quantitative benefits that are shown in this paper, MicroGrad is built in a modular manner with clear interface boundaries both internally as well as externally.
This allows it to be a promising springboard for wide future development - be it in terms of the use cases it can support, the evaluation platforms it can execute on, as well as running more optimum tuning algorithms.

For example, MicroGrad can seamlessly support other use cases like bottleneck analysis i.e. sweeping over a specified range of finer execution characteristics --such as cache miss rate-- and analyzing its bottle-necking impact on the overall processor execution. 
The framework also allows for experiments on native hardware and other forms of stress testing like voltage droops.
Thus, we envision that with future development, MicroGrad can accelerate the entire Innovate-Build-Analyze cycle as a whole, which is especially critical in the coming open-source hardware era.

\bibliographystyle{IEEEtranS}
\balance
\bibliography{refs}

\end{document}